\documentclass[a4paper,fleqn,usenatbib]{mnras}

\usepackage[T1]{fontenc}
\usepackage{ae,aecompl}

\usepackage{graphicx}   
\usepackage{amsmath}    
\usepackage{bm}       
\usepackage{amssymb}    
\usepackage{wasysym}  
\usepackage{txfonts}
\usepackage{etoolbox}
\makeatletter 
  \patchcmd{\NAT@citex}
    {\@citea\NAT@hyper@{%
      \NAT@nmfmt{\NAT@nm}%
      \hyper@natlinkbreak{\NAT@aysep\NAT@spacechar}{\@citeb\@extra@b@citeb}%
      \NAT@date}}
    {\@citea\NAT@nmfmt{\NAT@nm}%
    \NAT@aysep\NAT@spacechar\NAT@hyper@{\NAT@date}}{}{}

  \patchcmd{\NAT@citex}
    {\@citea\NAT@hyper@{%
      \NAT@nmfmt{\NAT@nm}%
      \hyper@natlinkbreak{\NAT@spacechar\NAT@@open\if*#1*\else#1\NAT@spacechar\fi}%
        {\@citeb\@extra@b@citeb}%
      \NAT@date}}
    {\@citea\NAT@nmfmt{\NAT@nm}%
    \NAT@spacechar\NAT@@open\if*#1*\else#1\NAT@spacechar\fi\NAT@hyper@{\NAT@date}}
    {}{}
\makeatother

\newcommand\Msun{\text{M}_{\astrosun}} 
\newcommand\colt{\textsc{colt}} 
\newcommand\HI{{H\,\textsc{i}}} 
\newcommand\HII{{H\,\textsc{ii}}} 
\newcommand\HeII{{He\,\textsc{ii}}} 

\def\app#1#2{%
  \mathrel{%
    \setbox0=\hbox{$#1\sim$}%
    \setbox2=\hbox{%
      \rlap{\hbox{$#1\propto$}}%
      \lower1.2\ht0\box0%
    }%
    \raise0.25\ht2\box2%
  }%
}
\def\approxprop{\mathpalette\app\relax}

\title[Radiative effects during the assembly of DCBHs]{Radiative effects during the assembly of direct collapse black holes}

\author[A.\ Smith et al.]{
  Aaron~Smith,$^1$\thanks{E-mail: \href{mailto:asmith@astro.as.utexas.edu}{asmith@astro.as.utexas.edu}}
  Fernando~Becerra,$^2$
  Volker~Bromm$^1$ and
  Lars~Hernquist$^2$
  \\
  $^1$Department of Astronomy, The University of Texas at Austin, Austin, TX 78712, USA \\
  $^2$Department of Astronomy, Harvard University, 60 Garden Street, Cambridge, MA 02138, USA
}

\date{Accepted 2017 August 1. Received 2017 July 25; in original form 2017 June 8}

\pubyear{2017}

\begin{document}
\label{firstpage}
\pagerange{\pageref{firstpage}--\pageref{lastpage}}
\maketitle

\begin{abstract}
  We perform a post-processing radiative feedback analysis on a 3D ab initio cosmological simulation of an atomic cooling halo under the direct collapse black hole (DCBH) scenario. We maintain the spatial resolution of the simulation by incorporating native ray-tracing on unstructured mesh data, including Monte Carlo Lyman~$\alpha$~(Ly$\alpha$) radiative transfer. DCBHs are born in gas-rich, metal-poor environments with the possibility of Compton-thick conditions, $N_\text{H} \gtrsim 10^{24}\,\text{cm}^{-2}$. Therefore, the surrounding gas is capable of experiencing the full impact of the bottled-up radiation pressure. In particular, we find that multiple scattering of Ly$\alpha$ photons provides an important source of mechanical feedback after the gas in the sub-parsec region becomes partially ionized, avoiding the bottleneck of destruction via the two-photon emission mechanism. We provide detailed discussion of the simulation environment, expansion of the ionization front, emission and escape of Ly$\alpha$ radiation, and Compton scattering. A sink particle prescription allows us to extract approximate limits on the post-formation evolution of the radiative feedback. Fully coupled Ly$\alpha$ radiation hydrodynamics will be crucial to consider in future DCBH simulations.
\end{abstract}



\section{Introduction}
\label{sec:introduction}
Quasars powered by black holes with masses up to $\sim 10^{10}\,\Msun$ have been observed out to a redshift of $z \sim 7$, presenting a timing problem for their growth by conventional accretion \citep{Fan_2006,Mortlock_2011,Wu_2015}. In the context of galaxy formation theory, the most plausible explanation for the origin of the first supermassive black holes (SMBHs) is either rapid seeding or unusually efficient growth \citep{Volonteri_2012,Johnson_Haardt_2016,Smith_AG_2017}. In particular, viable candidates include the direct collapse scenario which gives rise to $10^{4-6}\,\Msun$ seeds under specialized conditions in the early Universe \citep{Bromm_Loeb_2003,Begelman_2006,Lodato_2006,Regan_2009}, and hyper-Eddington accretion on to stellar remnant black holes \citep{Wyithe_Loeb_2012,Pacucci_2015,Inayoshi_2016}. Thereafter, episodic galaxy mergers and cold gas accretion from the cosmic web regulate the growth of nuclear black holes \citep{Li_2007,Mayer_2010,Smidt_2017}. The emergence of $M_\bullet \lesssim 10^6\,\Msun$ black holes within quasar progenitors at $z \gtrsim 10$ is beyond the reach of current telescopes, but should come within reach with the \textit{James Webb Space Telescope} (\textit{JWST}) \citep[e.g.][]{Natarajan_2017,Dayal_2017}. Indeed, different formation scenarios may have unique observational signatures and dynamical implications \citep{Gallerani_2017}. Therefore, our understanding of black hole assembly and growth in realistic settings can have a significant impact on models and simulations of the first galaxies \citep{Bromm_Yoshida_2011}. The first SMBHs can affect galaxy evolution through radiative feedback that drives winds, suppresses star formation, and contributes to the reionization of the Universe \citep{Loeb_Furlanetto_2013}.

Forming a direct collapse black hole (DCBH) requires primordial gas to collapse without fragmentation. This is achieved by strong non-ionizing Lyman--Werner radiation (LW; $h \nu = 11.2-13.6\,\text{eV}$) from neighbouring galaxies which photodissociates H$_2$ \citep{Agarwal_2012,Agarwal_2014,Regan_2017}. With only atomic hydrogen line cooling available, the gas follows nearly isothermal collapse once the virial temperature reaches $T_\text{vir} \sim 10^4$\,K \citep{Omukai_2001,Oh_Haiman_2002}. DCBHs are thus born in gas-rich environments with no previous star formation. The hydrogen column density when averaged over all directions is extremely high, $N_\text{H} \gtrsim 10^{24}\,\text{cm}^{-2}$, potentially leading to Compton-thick conditions \citep{Yue_2013}. Certainly, the initial assembly environment is self-absorbing to ionizing photons. However, the subsequent evolution remains an open question. In particular, what is the role of radiative feedback? After a certain time-scale metal enrichment from neighbouring galaxies and supernovae from concurrent star formation will alter the surroundings. Eventually, the black hole likely resembles that of typical active galactic nuclei (AGN), undergoing sub-Eddington accretion rates over multiple duty cycles \citep{Milosavljevic_2009}. However, is the dominant driver of the evolution of DCBHs related to internal or external factors?

A particularly interesting source of feedback in the DCBH scenario is Lyman~$\alpha$ (Ly$\alpha$) radiation pressure. As a strong resonant line of neutral hydrogen, Ly$\alpha$ photons undergo multiple scattering which may enhance the effective Ly$\alpha$ force by one or two orders of magnitude in optically thick regions \citep{Dijkstra_Loeb_2008,Dijkstra_Loeb_2009}. The first self-consistent Ly$\alpha$ radiation hydrodynamics (RHD) simulations were performed by \citet{Smith_CR7_2016,Smith_RHD_2017}, who coupled a Monte Carlo radiative transfer (MCRT) code with spherically symmetric Lagrangian frame hydrodynamics including ionizing radiation, non-equilibrium chemistry and cooling, and self-gravity. They found a dense shell-like outflow structure, formed in response to the central ionizing source, and that Ly$\alpha$ radiation pressure may have a significant dynamical impact on gas surrounding DCBHs. Finally, it has also been suggested that trapped Ly$\alpha$ cooling radiation may enhance the formation of DCBHs by facilitating the photodetachment of H$^-$ ions, precursors to H$_2$, during collapse \citep{Agarwal_2015,Johnson_Dijkstra_2017}. However, all Ly$\alpha$ feedback studies thus far have been limited to 1D geometries.

There is a good reason to believe that SMBH formation scenarios may be observationally testable with the capabilities of next-generation observatories. Indeed, already the luminous COSMOS redshift 7 (CR7) Ly$\alpha$ emitter at $z = 6.6$ has received substantial attention due to its exceptionally strong Ly$\alpha$ and possible \HeII\ 1640~\AA\ emission with no detection of metal lines from the UV to the near-infrared within instrumental sensitivity \citep{Matthee_2015,Sobral_2015}. As a result, CR7 has served as a prominent case study for young primordial starburst and DCBH models \citep{Pallottini_2015,Agarwal_2016,Dijkstra_DCBH_2016,Hartwig_2016,Smidt_CR7_2016,Smith_CR7_2016,Visbal_CR7_2016}. However, the debate regarding the origin of CR7 remains ongoing, especially after deeper broad-band observations are consistent with a low-mass, narrow-line AGN or stellar models with massive, low-metallicity binaries \citep{Bowler_CR7_2017}. Furthermore, recent reexamination of the original spectroscopic data lowers the confidence-level of the \HeII\ line detection to $< 2 \sigma$ \citep{Shibuya_2017}. Deeper spectroscopy of CR7 and future candidates will be necessary to convincingly discriminate between a black hole seeded via direct collapse and other more standard interpretations \citep{Agarwal_2017,Pacucci_CR7_2017}. Still, other observational signatures of DCBHs may be manifest if their spectral energy distributions (SEDs) exhibit the imprint of Compton-thickness \citep{Pacucci_MBH_Spectra_2015}. For example, DCBHs may be responsible for the measured correlations between the cosmic infrared and X-ray backgrounds, i.e. the source-subtracted fluctuations after accounting for foreground stars and galaxies \citep{Kashlinsky_2005,Cappelluti_2013,Helgason_2016}. The collection of several independent signatures with next-generation observatories will clarify and constrain our understanding of the first SMBHs.

The primary focus of this paper is to investigate the effects of radiative feedback during DCBH assembly by carrying out a post-processing analysis of the high-resolution ab initio cosmological simulation of \citet{Becerra_2017}. For this exploratory study, the radiation remains decoupled from the hydrodynamical evolution of the gas. However, we fully explore the impact of Ly$\alpha$ radiation pressure in a realistic 3D DCBH environment, which is therefore complementary to the work of \citet{Smith_RHD_2017}. This is important because asymmetries in the gas distribution introduce a directional dependence of the Ly$\alpha$ spectrum and escape properties \citep[e.g.][]{Behrens_2014,Zheng_Wallace_2014,Smith_2015}. Throughout this work, we use the terminology of pre- and post-formation to consider environmental changes due to the buildup of the massive object \citep{Becerra_2017,Woods_2017}. This paper is organized as follows. In Section~\ref{sec:simulation}, we discuss details of the simulation setup as well as the physical properties of the host galaxy. In Section~\ref{sec:ion}, we argue that the buildup and eventual breakout of ionizing photons represents a critical transition for the DCBH. In Section~\ref{sec:Lya}, we present the Ly$\alpha$ radiative transfer calculations and discuss the potential impact in relation to other forces acting on the gas. In Section~\ref{sec:compton}, we evaluate the role of Compton scattering in the sub-parsec region around the black hole. Finally, in Section~\ref{sec:conc}, we explore the implications of our work.

\section{Simulation}
\label{sec:simulation}
%

  \begin{figure*}
    \centering
    \includegraphics[width=\textwidth]{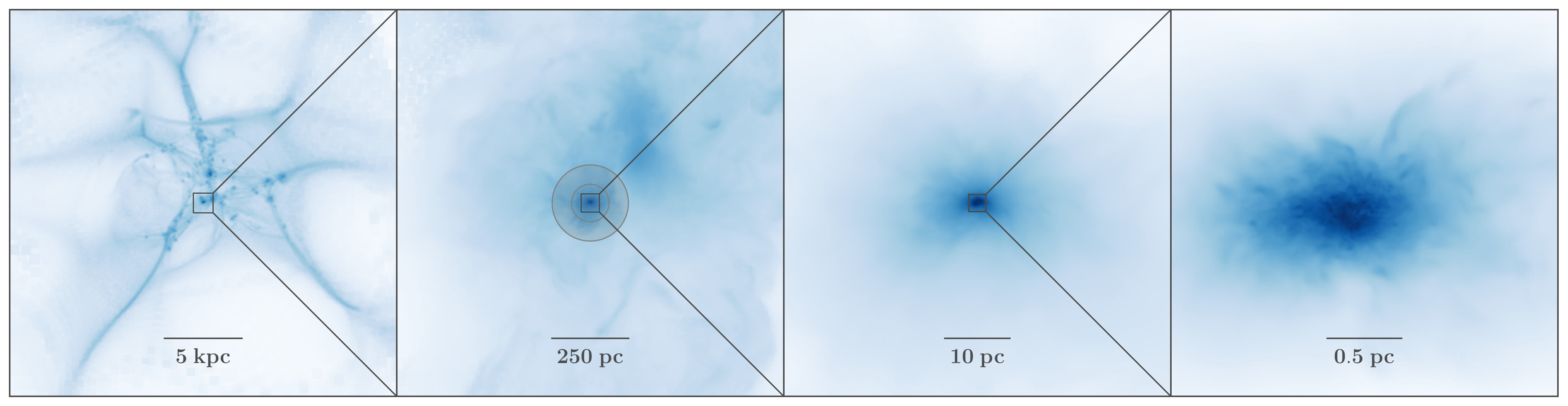}
    \caption{\protect\input{figures/density/caption}}
    \label{fig:density}
  \end{figure*}

\subsection{DCBH Formation}
We perform a high-resolution ab initio cosmological simulation of the collapse of an atomic cooling halo. Using the approach described in \citet{Becerra_2015} and \citet{Becerra_2017}, we follow the evolution of the halo from cosmological initial conditions at redshift $z = 99$ until its collapse at $z \approx 11.6$. Fig.~\ref{fig:density} shows the projected gas density of the central 25~kpc, 1.25~kpc, 50~pc and 2.5~pc from left to right, which illustrates the morphology of the halo from the large-scale environment down to the central gas cloud. We emulate the formation of a DCBH by inserting a sink particle once the highest density cell reaches a threshold number density of $n_\text{th} \simeq 10^8\,\text{cm}^{-3}$. The accretion radius of this particle is set to $R_\text{acc} = 1\,\text{pc}$ \citep{Becerra_2017} in such a way that the initial mass of the sink particle is $M_\bullet \approx 8.8 \times 10^4\,\Msun$. Thus, the post-formation scenario should be interpreted as an aggressive extrapolation of the DCBH environment without fully resolving the formation and growth of the massive object in the hydrodynamical simulation.

  \begin{figure}
    \centering
    \includegraphics[width=\columnwidth]{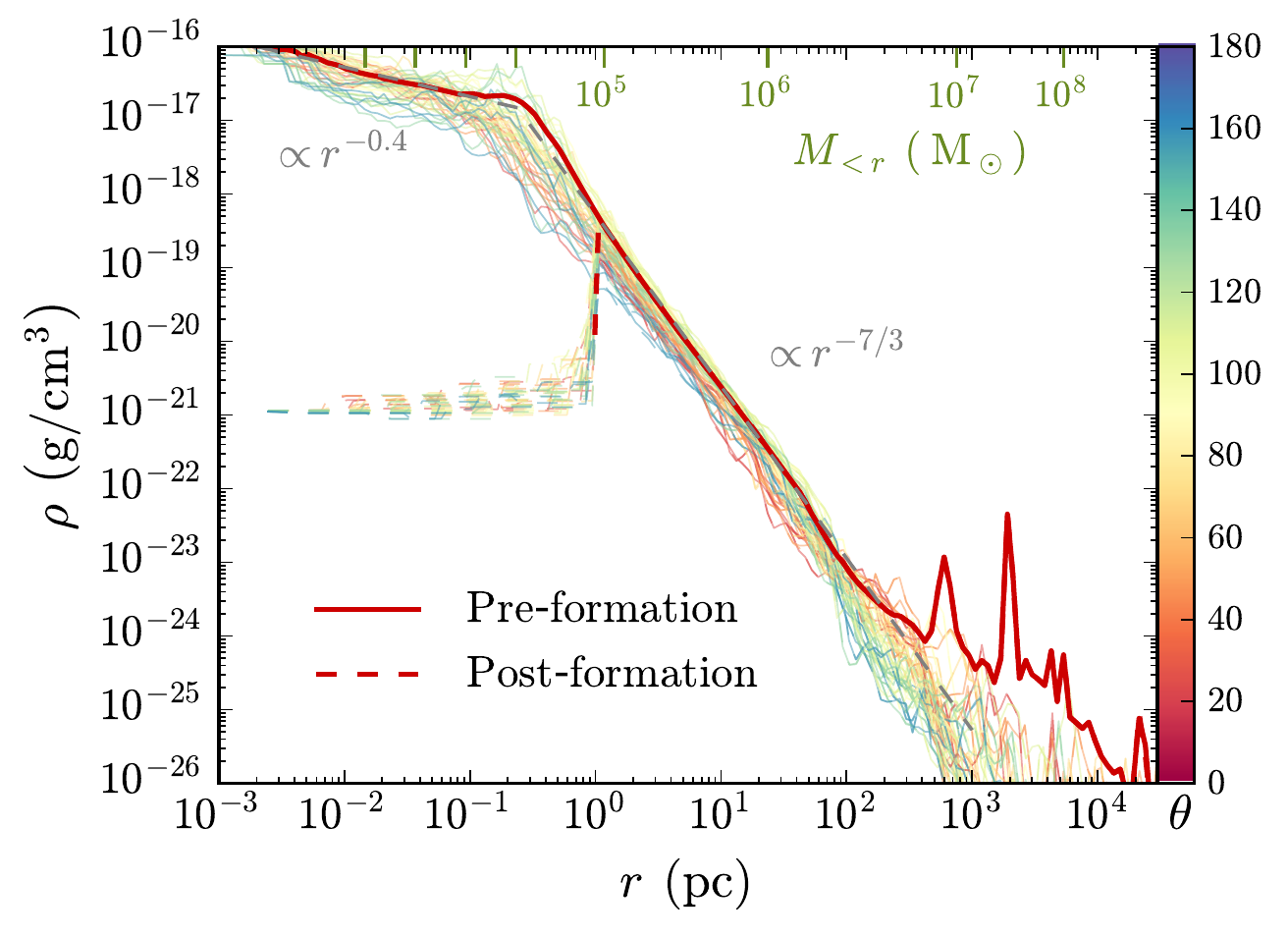}
    \caption{\protect\input{figures/rho/caption}}
    \label{fig:rho}
  \end{figure}

  \begin{figure}
    \centering
    \includegraphics[width=\columnwidth]{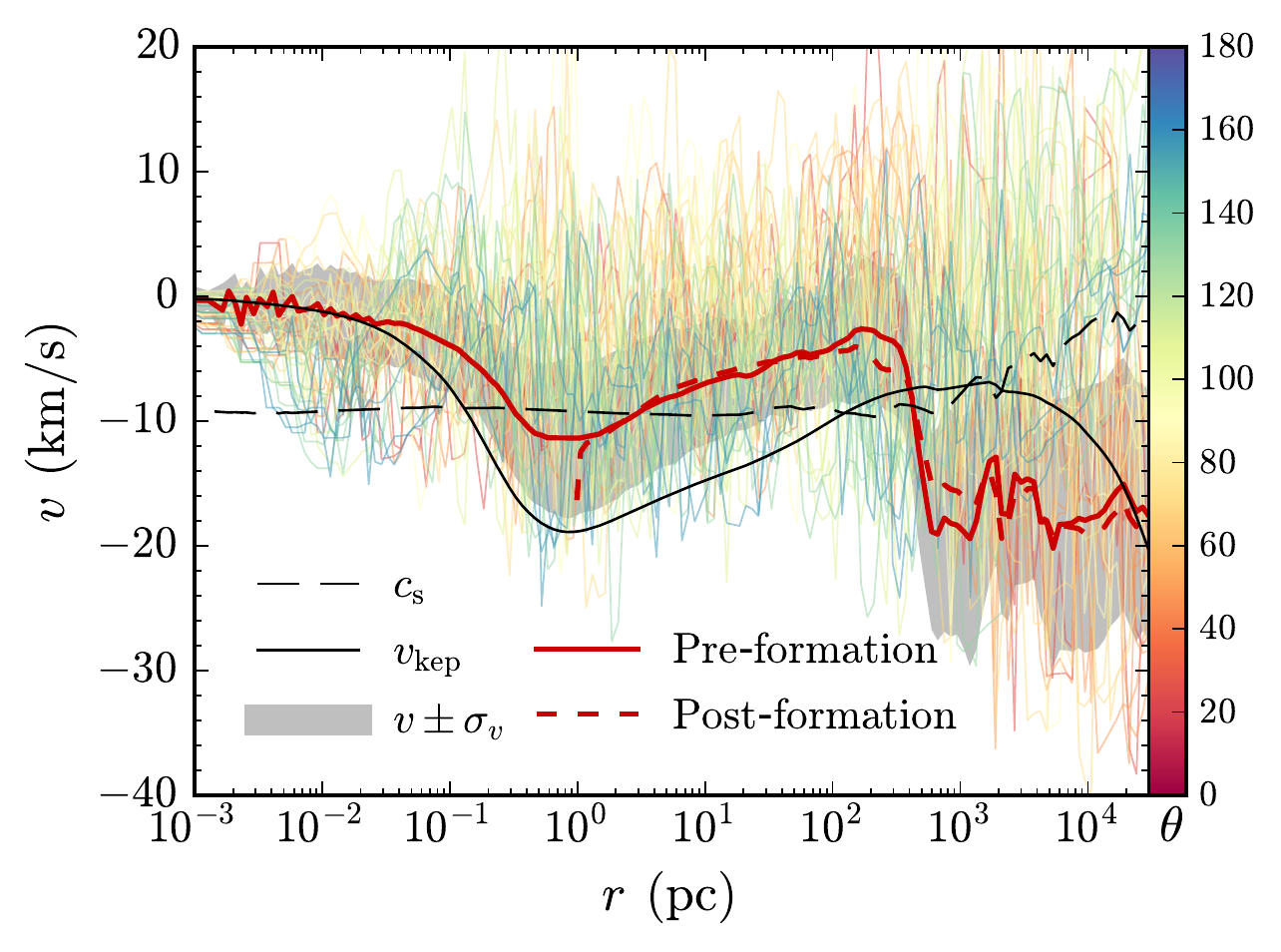}
    \caption{\protect\input{figures/v/caption}}
    \label{fig:v}
  \end{figure}

\subsection{Insights from radial profiles}
Throughout the initial collapse the halo structure is well approximated by ellipsoidal collapse models. We therefore explore radial profiles of various physical quantities to extract information about the galactic environment. The density is illustrated in Fig.~\ref{fig:rho}, which is reasonably approximated by a broken power law over the range $r \in (10^{-3}, 10^3)$~pc. The break radius is due to the choice of the density resolution threshold, or equivalently, the limited resolution implies the pre-formation profile corresponds to roughly a dynamical time, as evaluated at the maximum resolved density, of $\sim 10$\,kyr prior to the formation of the protostar. If the evolution proceeds under self-similar, isothermal collapse then the break in the profile will shift to smaller scales, eventually reaching the radius of the protostar \citep{Abel_2002,Becerra_2015}. We note that secondary infall and accretion results in a density distribution that is steeper than the $\rho \propto r^{-2}$ profile produced by violent relaxation. We find a power-law scaling of $\rho \approxprop r^{-7/3}$ in good agreement with \citet{Bertschinger_1985} where $\rho \propto r^{-9/4}$ and \citet{Wise_2008} where $\rho \propto r^{-12/5}$, except in the core where the slope flattens off to $\rho \propto r^{-0.4}$ around $r \approx 0.3$~pc. During the initial collapse and formation, the gas remains neutral with only a small abundance of free protons. Specifically, the ionization fraction, $x_\text{\HII} \equiv n_\text{\HII} / n_\text{H}$, is typically of order $10^{-5}$ to $10^{-3}$ depending on the density, such that $n_\text{\HII} \sim 0.1\,\text{cm}^{-3}\,(r/10\,\text{pc})^{-5/3}$. As expected for direct collapse, the central region has access to at least $10^5\,\Msun$ of gas within a radius of $\approx 1.3$~pc. More broadly, the enclosed baryonic mass is $M_{<r} \approx 4 \pi \int_0^r \rho(r) r^2 \text{d}r$ (for 1D profiles), calculated explicitly as $M_{<r} \equiv \sum_{<r} \rho_i V_i$ where the sum is over all Voronoi cells within a radius $r$, and the subscript denotes cell quantities for density~$\rho_i$ and volume~$V_i$. For convenience and completeness, in Table~\ref{tab:scalings} we provide radial scaling relations for several relevant quantities, calculated as mass-weighted averages within shell volumes, i.e.
\begin{equation} \label{eq:shell_average}
  \langle q \rangle_\text{shell} \equiv \frac{\int q \rho \, \text{d}V}{\int \rho \, \text{d}V} \approx \frac{\sum q_i \rho_i V_i}{\sum \rho_i V_i} \, ,
\end{equation}
where the discretized version sums over all cells within each shell. Similarly, we obtain mass-weighted line of sight averages with
\begin{equation} \label{eq:LOS_average}
  \langle q \rangle_\text{LOS} \equiv \frac{\int q \rho \, \text{d}\ell}{\int \rho \, \text{d}\ell} \approx \frac{\sum q_i \rho_i \Delta\ell_i}{\sum \rho_i \Delta\ell_i} \, ,
\end{equation}
where the integral is along radial rays within each shell and the summation is the discretized version calculated by ray tracing.

\begin{table}
  \caption{Radial scaling relations for various physical quantities calculated as mass weighted averages. Most of the profiles are reasonably approximated by a broken power law separated into regions~\textsc{i}~and~\textsc{ii} as $r \in (10^{-3}, r_{\textsc{i}\text{-}\textsc{ii}})$~pc and $r \in (r_{\textsc{i}\text{-}\textsc{ii}}, 10^3)$~pc, respectively, where $r_{\textsc{i}\text{-}\textsc{ii}}$ is the radius connecting the regions. For convenience, the scalings are normalized according to values representative of the centre of each region, i.e. $r_\textsc{i} \equiv r / (0.05~\text{pc})$ and $r_\textsc{ii} \equiv r / (10~\text{pc})$. All relations are approximate fits to simulation data based on shell and line-of-sight averages described by equations~(\ref{eq:shell_average})~and~(\ref{eq:LOS_average}). See the text for definitions and descriptions of all quantities.}
  \label{tab:scalings}
  \begin{tabular}{@{} lr ccc @{}}
    \hline
    Quantity & Units & Region \textsc{i} & Region \textsc{ii} & \hspace{-.4cm} $r_{\textsc{i}\text{-}\textsc{ii}}$\;[pc] \\
    \hline
    \vspace{.05cm}
    $\rho$ & \hspace{-.5cm} [g\,cm$^{-3}$] & \hspace{-.15cm} $2.78 \times 10^{-17} r_\textsc{i}^{-0.4}$ & \hspace{-.25cm} $2.46 \times 10^{-21} r_\textsc{ii}^{-7/3}$ & $0.24$ \\
    $n_\text{H}$ & \hspace{-.5cm} [cm$^{-3}$] & $1.26 \times 10^7 \, r_\textsc{i}^{-0.4}$ & $1.12 \times 10^3 \, r_\textsc{ii}^{-7/3}$ & $0.24$ \\
    $x_\text{\HII}$ & \hspace{-.5cm} [$1$] & $1.15 \times 10^{-5} \, r_\textsc{i}^{-0.3}$ & $1.15 \times 10^{-4} \, r_\textsc{ii}^{2/3}$ & $0.18$ \\
    $M_{<r}$ & \hspace{-.5cm} [$\Msun$] & $214 \; r_\textsc{i}^{2.6}$ & $5.10 \times 10^5 \, r_\textsc{ii}^{2/3}$ & $0.45$ \\
    $\dot{M}$ & \hspace{-.5cm} [$\Msun$/yr] & $0.0325 \; r_\textsc{i}^{2.3}$ & $0.327 \; r_\textsc{ii}^{-2/3}$ & $0.36$ \\
    $v_r$ & \hspace{-.5cm} [km\,s$^{-1}$] & $-2.47 \; r_\textsc{i}^{0.7}$ & $-6.99 \; r_\textsc{ii}^{-1/4}$ & $0.6$ \\
    $\sigma_v$ & \hspace{-.5cm} [km\,s$^{-1}$] & $3.83 \; r_\textsc{i}^{0.18}$ & $4.94 \; r_\textsc{ii}^{-0.1}$ & $0.82$ \\
    $v_\text{kep}$ & \hspace{-.5cm} [km\,s$^{-1}$] & $4.29 \; r_\textsc{i}^{0.8}$ & $14.8 \; r_\textsc{ii}^{-1/6}$ & $0.45$ \\
    $j_\text{kep}$ & \hspace{-.5cm} [pc\,km\,s$^{-1}$] & $0.215 \; r_\textsc{i}^{1.8}$ & $148 \; r_\textsc{ii}^{5/6}$ & $0.45$ \\
    $|j|/j_\text{kep}$ & \hspace{-.5cm} [1] & $< 10$ & $\approx 0.4-0.6$ & $0.01$ \\
    $T$ & \hspace{-.5cm} [K] & $\approx 7000 - 8000$ & $\approx 7000 - 8000$ & -- \\
    $c_\text{s}$ & \hspace{-.5cm} [km\,s$^{-1}$] & $\approx 9-9.5$ & $\approx 9-9.5$ & -- \\
    $t_\text{ff}$ & \hspace{-.5cm} [Myr] & $0.0127 \; r_\textsc{i}^{0.2}$ & $0.733 \; r_\textsc{ii}^{7/6}$ & $0.45$ \\
    $\lambda_\text{J}$ & \hspace{-.5cm} [pc] & $0.181 \; r_\textsc{i}^{0.2}$ & $11.2 \; r_\textsc{ii}^{7/6}$ & $0.42$ \\
    $\lambda_\text{J}/2r$ & \hspace{-.5cm} [1] & $3.61 \; r_\textsc{i}^{-0.8}$ & $1.12 \; r_\textsc{ii}^{1/6}$ & $0.42$ \\
    $a_\text{grav}$ & \hspace{-.875cm} [km\,s$^{-1}$Myr$^{-1}$] & $413 \; r_\textsc{i}^{0.6}$ & $38.1 \; r_\textsc{ii}^{-1.1}$ & $0.38$ \\
    $a_\text{P}$ & \hspace{-.875cm} [km\,s$^{-1}$Myr$^{-1}$] & $536 \; r_\textsc{i}^{-1}$ & $12.5 \; r_\textsc{ii}^{-1}$ & $0.3$ \\
    $a_\text{ram}$ & \hspace{-.875cm} [km\,s$^{-1}$Myr$^{-1}$] & $92.2 \; r_\textsc{i}^{0.4}$ & $1.08 \; r_\textsc{ii}^{-3/2}$ & $0.32$ \\
    $a_\text{P}/a_\text{grav}$ & \hspace{-.5cm} [1] & $1.30 \; r_\textsc{i}^{-1.6}$ & $0.328 \; r_\textsc{ii}^{0.1}$ & $0.15$ \\
    \hline
  \end{tabular}
\end{table}

As seen in Fig.~\ref{fig:v}, the gas is experiencing both cosmological inflow throughout the halo and collapse within the central region. To quantify this, we consider the mass accretion rate $\dot{M} \equiv -4 \pi r^2 \rho v_r$, which has a maximum value of $2.35~\Msun\,\text{yr}^{-1}$ at $0.465$~pc just outside the core radius. In Fig.~\ref{fig:v}, we also show the mass-weighted velocity along multiple sightlines and find a considerable scatter across different directions, as quantified by the velocity dispersion (grey region) in each shell as $\sigma_v \equiv ( \langle v^2 \rangle - \langle v \rangle^2 )^{1/2}$. Angular momentum does not appear to inhibit the inflow in this case. In particular, we calculate the Keplerian or rotational velocity as $v_\text{kep} \equiv \sqrt{G M_{<r} / r}$ (solid black curve) and find that $|v_r| \approx v_\text{kep} / 2$ within the virial radius and $|v_r| \approx 2\,v_\text{kep}$ beyond, the latter is a reflection of cold gas inflow along the filaments. The associated Keplerian angular momentum is $j_\text{kep} = r\,v_\text{kep}$ and calculating the magnitude of the mass-weighted specific angular momentum demonstrates that $|j| / j_\text{kep} \approx 0.4-0.6$ throughout the halo except in the immediate vicinity of the collapsing region ($\lesssim 0.01$~pc) where the role of angular momentum is less certain. In Fig.~\ref{fig:v}, we also show the sound speed (dashed black curve) defined as $c_\text{s} \equiv \sqrt{\gamma k_\text{B} T / \mu}$, where in primordial gas the adiabatic index is $\gamma = 5/3$ and the mean molecular weight is $\mu \approx 1.22\,m_\text{H}$. However, atomic line cooling efficiently regulates the temperature to $T \approx 7000-8000$~K, so the sound speed is also roughly constant at $c_\text{s} \approx 9-9.5$~km\,s$^{-1}$ throughout the halo.

Dynamical quantities are also of interest. We find short free-fall times within the core ($\approx 5-25$\,kyr), rapidly rising beyond, such that $t_\text{ff} \approx \{15, 60, 775\}$~kyr at $r = \{0.1, 1, 10\}$~pc, respectively. Here, $t_\text{ff} \equiv \sqrt{3 \pi / 32 G \bar{\rho}}$, where $\bar{\rho} \equiv 3 M_{<r} / 4 \pi r^3$ is the average density within a given radius. To quantify the stability of the system we consider the Jeans length $\lambda_\text{J} \equiv \sqrt{15 k_\text{B} T / 4 \pi G \mu \bar{\rho}} \sim t_\text{ff\,} c_\text{s}$. Interestingly, the ratio of the Jeans length to the diameter has a minimum value just outside the core radius, i.e. min$(\lambda_\text{J}/2r) = 0.84$ at $r = 0.83$~pc, indicating that the break in the roughly isothermal scaling may be related to this instability length-scale. However, the break may also partially be a result of limited numerical resolution in the regions of highest density. Finally, outside the core both the gravitational and gas pressure acceleration components scale inversely with radius, $a \approxprop r^{-1}$. This is as expected for near-isothermal profiles in which the density profile $\rho \propto r^{-2}$ gives an enclosed mass of $M_{<r} \propto r$ so the acceleration is $a_\text{grav} \propto M_{<r}/r^2 \approx r^{-1}$. Furthermore, the gas pressure is roughly proportional to the density, i.e. thermal pressure $\sim \rho\,k_\text{B\,}T$ with constant temperature or ram pressure $\sim \rho\,\sigma^2$ with constant velocity dispersion, so the force is $a_\text{P} \propto \text{d}\log\rho/\text{d}r \approx r^{-1}$. The radial deceleration due to ram pressure, defined by $a_\text{ram} \equiv v_r \frac{\text{d}v_r}{\text{d}r}$, exhibits a steeper radial profile of $\propto r^{-3/2}$, but is about an order of magnitude smaller than the other forces. In general, the accelerations can have large variations across different sightlines, but gravity dominates the overall gas dynamics during the collapse stage.

\section{Ionizing radiation}
\label{sec:ion}
%

  \begin{figure}
    \centering
    \includegraphics[width=\columnwidth]{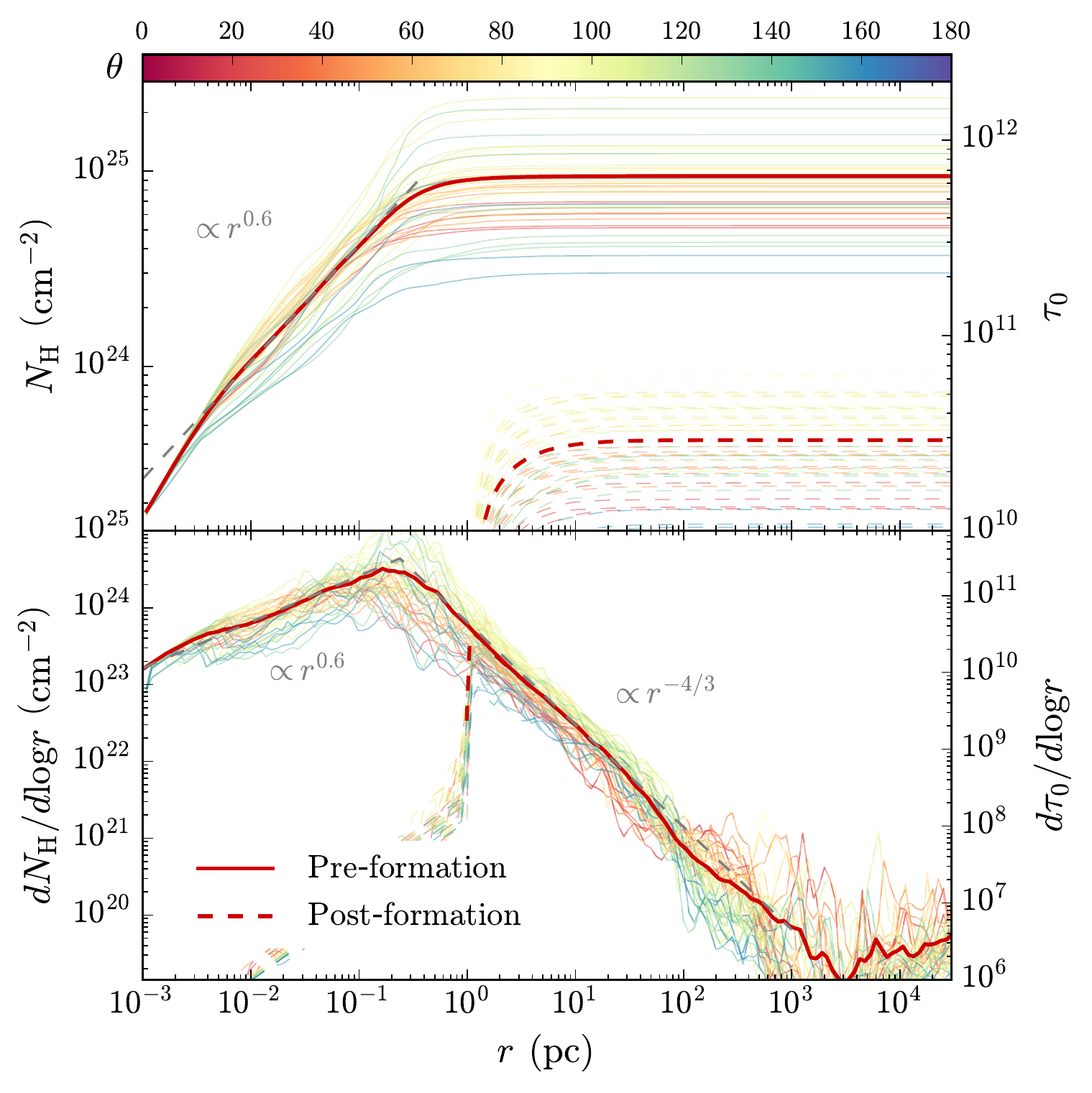}
    \caption{\protect\input{figures/N_H/caption}}
    \label{fig:N_H}
  \end{figure}

\subsection{Neutral hydrogen column density}
During the collapse phase the central region accumulates a significant amount of dense neutral gas. As seen in Fig.~\ref{fig:N_H}, the dominant contribution to the radially-integrated opacity is from the core, where the hydrogen column density is $N_\text{H}(r) \equiv \int_0^r n_\text{H}\,\text{d}\ell \sim 10^{25}$\,cm$^{-2}$ and the optical depth for ionizing radiation is $\tau_\text{ion} \approx N_\text{H} \sigma_\text{ion} \gtrsim 10^7$ assuming $\sigma_\text{ion} \gtrsim 10^{-18}~\text{cm}^2$ near the 13.6\,eV hydrogen ionization threshold. The variation along sightlines is roughly one order of magnitude with the minimum columns oriented along the poles of the angular momentum axis. Fig.~\ref{fig:N_H} also shows the optical depth for Ly$\alpha$ resonant scattering at line centre, which is $\tau_0 \approx N_\text{H\,} \sigma_0$ in isothermal gas with a Ly$\alpha$ cross-section at line centre of $\sigma_0 = 5.9 \times 10^{-14}\,\text{cm}^2\,T_4^{-1/2}$, where $T_4 \equiv T / (10^4\,\text{K})$.

\subsection{Expansion of the ionization front}
During the formation of the central supermassive object, the high density implies that the ionization front is bounded within the dense sub-parsec core. However, as the black hole grows, at least three effects are likely to allow the ionization front to eventually break out of the core within a relatively short time. First, the number of ionizing photons increases rapidly; secondly, accretion removes gas efficiently enough to lower the ionization rate threshold to ionize the core; and thirdly, dramatic photoheating of the gas leads to rapid expansion, facilitating ionization by lowering the density and recombination rate. Once the ionization front has extended beyond the core into the $\approxprop r^{-7/3}$ profile, it quickly outruns the rest of the halo in approximately the light crossing time. We estimate the breakout to occur on the order of $\lesssim 10^5$\,yr, which corresponds to the sound crossing time for the core.

  \begin{figure}
    \centering
    \includegraphics[width=\columnwidth]{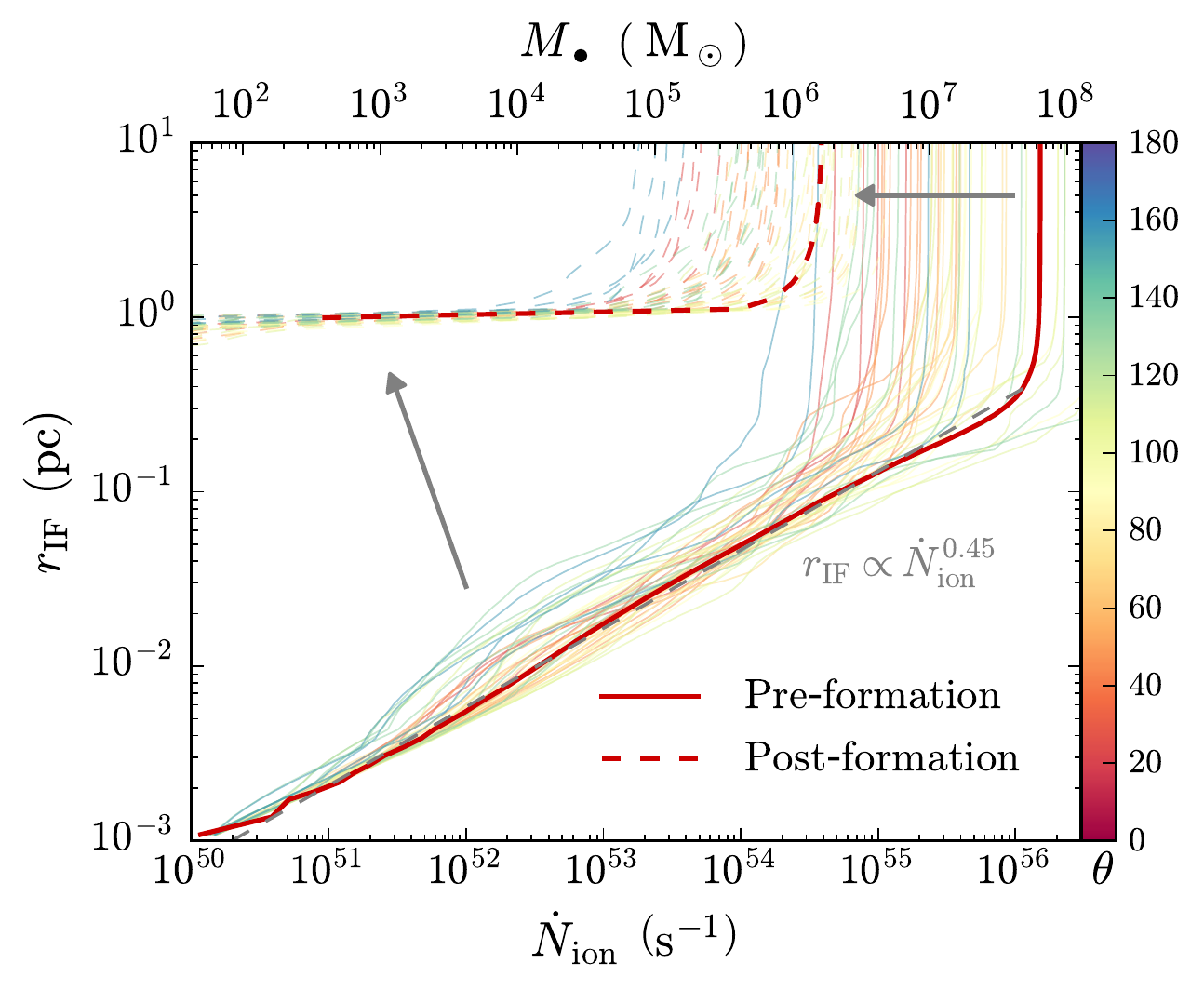}
    \caption{\protect\input{figures/r_IF/caption}}
    \label{fig:r_IF}
  \end{figure}

To analyse the expansion of the ionization front we consider an equilibrium model in which recombinations balance ionizations within a Str{\"o}mgren sphere. The number of ionizing photons per unit time emitted from the source needed to produce an ionized region of radius $r_\text{IF}$ based on the density in the simulation is
\begin{equation} \label{eq:Ndot_ion_IF}
  \dot{N}_\text{ion} \approx 4 \pi \int_0^{r_\text{IF}} \alpha_\text{B} n_\text{H}^2 r^2 \text{d}r \, ,
\end{equation}
where $\alpha_\text{B} = 2.59 \times 10^{-13}\,T_4^{-0.7}\,\text{cm}^3\,\text{s}^{-1}$ is the effective Case B recombination coefficient \citep{Osterbrock_2006}. This analysis is valid until dynamical processes kick in (e.g. accretion and radiative feedback), because the recombination time, $t_\text{rec} \sim 1/n_\text{H} \alpha_\text{B} \lesssim 1\,\text{yr}$, is much shorter than the sound crossing time, $t_\text{s} \equiv r / c_\text{s} \approx 10^{5}\,\text{yr}\,(r/1\,\text{pc})$. The main goal of this paper is to determine the impact of Ly$\alpha$ radiation pressure in the vicinity of a DCBH. However, the development of the \HII\ region will significantly modify the Ly$\alpha$ radiative transfer. Therefore, we consider how the full 3D ionization structure evolves in time. Even though the density profile is roughly spherically symmetric, the column density can differ by an order of magnitude between sightlines (compare Figs~\ref{fig:density}, \ref{fig:rho} and \ref{fig:N_H}). Fig.~\ref{fig:r_IF} demonstrates this effect by exhibiting the extent of the ionization front as a function of source ionization rate based on equation~(\ref{eq:Ndot_ion_IF}). The rate threshold to blow pockets out of the core is one to two orders of magnitude lower in the polar directions. Ultimately, the threshold is overcome for a significant solid angle as the accretion and intensifying source lead to an evolving system, as indicated by the arrows in Fig.~\ref{fig:r_IF}. The rate of ionizing photons is proportional to the black hole mass, although the details depend on the specific SED. To simplify the calculation, we assume a blackbody source emitting at the Eddington luminosity, $L_\text{Edd} = 4 \pi G M_\bullet m_\text{H} c / \sigma_\text{T}$, with an effective temperature of $T_\text{eff} = 10^5\,\text{K}$. Thus,
\begin{align} \label{eq:Ndot_ion_BB}
  \dot{N}_\text{ion} &= \frac{\pi L_\text{Edd}}{\sigma_\text{SB} T_\text{eff}^4} \int_{\nu_\text{min}}^\infty \frac{B_\nu}{h \nu} \text{d}\nu \notag \\
    &\approx 2.37 \times 10^{54}\,\text{s}^{-1}\,\left( \frac{M_\bullet}{10^6\,\Msun} \right) \, ,
\end{align}
where $\sigma_\text{SB}$ is the Stefan--Boltzmann constant, $B_\nu$ the Planck function, and $h \nu_\text{min} = 13.6\,\text{eV}$. However, a harder (softer) SED or more (less) efficient source produces more (fewer) ionizing photons per unit time, implying an uncertainty in the constant of proportionality of at least a factor of a few \citep{Bolton_2010,Yue_2013}.

Finally, we briefly explain why the halo is ionized so quickly once the ionization rate threshold is reached. We first approximate equation~(\ref{eq:Ndot_ion_IF}) by $\dot{N}_\text{ion} \propto n_\text{H,rms}^2 r_\text{IF}^3$, where the rms number density within the ionized region is $n_\text{H,rms} \equiv (3\,r_\text{IF}^{-3} \int_0^{r_\text{IF}} n_\text{H\,}^2 r^2\,\text{d}r)^{1/2}$. In our case the power law in the core is observed to be $n_\text{H} \propto r^{-0.4}$, which is preserved when calculating the rms density. This explains the origin of the scaling in the central region of Fig.~\ref{fig:r_IF}, where we find $\dot{N}_\text{ion} \propto (r_\text{IF}^{-0.4})^2 r_\text{IF}^3 = r_\text{IF}^{2.2}$, or an inverted relation of $r_\text{IF} \propto \dot{N}_\text{ion}^{0.45}$. Outside the core the profile falls off rapidly and the rms density integral is dominated by the contribution from the core, i.e. $\int_0^{r_\text{IF}} n_\text{H\,}^2 r^2\,\text{d}r \rightarrow \text{constant}$. Therefore, the rms scaling becomes $n_\text{H,rms} \propto r^{-3/2}$ and the required rate of ionizing photons saturates outside the core because $\dot{N}_\text{ion} \propto n_\text{H,rms}^2 r_\text{IF}^3 \simeq \text{const}$.

\section{Lyman-alpha trapping}
\label{sec:Lya}

\subsection{Sub-parsec optical depth}
During DCBH assembly the optical depth for Ly$\alpha$ photons at line centre can be significantly higher than what is typically described as extremely optically thick, conventionally defined by the condition $a\tau_0 \gtrsim 1000$ or equivalently $\tau_0 \gtrsim 2 \times 10^6\,T_4^{1/2}$ \citep{Adams_1972}. Here, the damping parameter $a \equiv \Delta \nu_\text{L} / 2 \Delta \nu_\text{D}$, is half the ratio of the natural line width to the thermally broadened Doppler width. To distinguish between the broad range of environments we term even higher opacity regimes as `hyper-extreme', as is the case for DCBH formation where $\tau_0 \sim 10^{12}$ (see Fig.~\ref{fig:N_H}). However, the escape of Ly$\alpha$ photons at hyper-extreme optical depths will be regulated by photoionization, two-photon emission, velocity gradients, dust absorption and 3D effects such as gas clumping, rotation, filamentary structure, holes or anisotropic emission. These effects are not independent of each other. For example, even at high densities collisional de-excitation becomes inefficient if the average number of scatterings remains lower than a certain threshold \citep{Dijkstra_DCBH_2016}. Which effects are most important to accurately model?

We first consider the nature of Ly$\alpha$ resonant scattering in extremely opaque media, which can be thought of as a diffusion process in both frequency and space \citep{Adams_1972,Harrington_1973,Neufeld_1990}. The frequency diffusion accounts for partially coherent scattering in the wings of the Ly$\alpha$ profile \citep{Unno_1952,Hummer_1962}, and reduces the mean number of scatterings from the strong $N_\text{scat} \propto \tau^2$ dependence under a spatial random walk to a nearly linear scaling $N_\text{scat} \propto \tau_0$ \citep{Osterbrock_1962,Adams_1972}. However, at these opacities the mean number of successive wing scatterings just before escape may still be described as a diffusion process at the effective wing optical depth, i.e. $N_\text{scat,w} = \tau_\text{w}^2 \approx (a \tau_0)^{2/3} / \upi \approx 19.25\,(\tau_0/10^6)^{2/3}\,T_4^{-1/3}$ \citep{Ahn_2002}, emphasizing the role of excursions to the wing in facilitating the propagation of Ly$\alpha$ photons. For additional details and discussion of Ly$\alpha$ radiative transfer see the review by \citet{Dijkstra_2014}.

\subsection{Two-photon decay}
\subsubsection{Number of scattering events}
In this paper we adopt a value of $N_\text{scat} \approx 0.6\,\tau_0$ based on numerical calculations in which Ly$\alpha$ photons are emitted in the centre of a uniform, static sphere \citep{Dijkstra_2006}. This scenario provides a conservative upper limit on $N_\text{scat}$ but should be close to the actual number for the following reasons. First, although the gas cloud is not static the infall velocity is comparable to the thermal velocity of the gas, and therefore only decreases $N_\text{scat}$ by a small amount \citep{Bonilha_1979,Dijkstra_3cm_2016}. Additionally, it has been shown that $N_\text{scat}$ is independent of bulk velocity for uniform spheres undergoing solid-body rotation up to $300\,\text{km\,s}^{-1}$ \citep{Garavito-Camargo_2014}. Secondly, in reality the Ly$\alpha$ radiation is not centrally emitted, either due to radial dependence or offset with respect to the centre of mass of the cloud. However, this too has been shown to have only a minor effect on $N_\text{scat}$, reducing the estimate by a factor of approximately $1.4-2$ for homogeneously distributed sources \citep{Harrington_1973,Garavito-Camargo_2014}. Thirdly, 3D inhomogeneities may also lower the number of scatterings. Indeed, the column density may differ by an order of magnitude across sightlines (see Fig.~\ref{fig:N_H}). However, translating this to $N_\text{scat}$ is non-trivial as multiple scattering tends to isotropize the radiation field. Even if Ly$\alpha$ photons are efficiently channeled towards low column pathways, in the DCBH scenario fragmentation is suppressed and the gas distribution remains fairly smooth, unlike multiphase models with clumpy sub-structures \citep{Neufeld_1991,Gronke_2017}.

\subsubsection{Escape fraction with de-excitation}
We now estimate the impact of collisional de-excitation in the simulation. During Ly$\alpha$ scattering interactions with nearby protons can induce transitions of the form $2p \rightarrow 2s$. Once in the $2s$-state, the excited atom returns to the ground state via two-photon decay. The probability that a Ly$\alpha$ photon is eliminated in this process per scattering event is
\begin{equation} \label{eq:p_dest}
  p_\text{dest} = \frac{n_p C_{ps}}{n_p C_{ps} + A_\alpha} 
                \approx 2.83 \times 10^{-7} \, \left(\frac{n_p}{10^6\,\text{cm}^{-3}}\right) \, ,
\end{equation}
where $n_p$ is the number density of free protons, $A_\alpha = 6.25 \times 10^8\,\text{s}^{-1}$ is the Einstein-A coefficient of the Ly$\alpha$ transition and $C_{ps} = 1.77 \times 10^{-4}\,\text{cm}^3\,\text{s}^{-1}$ is the collision strength, which has weak dependence on temperature \citep{Seaton_1955,Dennison_2005}. The fraction of Ly$\alpha$ photons that is \textit{not} destroyed after $N_\text{scat}$ scattering events at constant density and temperature is \citep{Dijkstra_DCBH_2016}
\begin{equation}
  f_\text{esc}^{2\gamma} = [1 - p_\text{dest}]^{N_\text{scat}} \approx 1 - N_\text{scat} p_\text{dest} \, ,
\end{equation}
where the second expression is valid when $p_\text{dest} \ll 1$. If we ignore helium ionizations then the free proton number density is $n_p \approx x_e n_\text{H}$, where $x_e$ denotes the ionized fraction. Additionally, a static uniform sphere gives $N_\text{scat} \approx 0.6\,\tau_0 \approx 3.28 \times 10^6\,T_4^{-1/2}\left(\frac{1-x_e}{10^{-4}}\right) \left(\frac{n_\text{H}}{10^6\,\text{cm}^{-3}}\right)\,\left(\frac{R}{0.3\,\text{pc}}\right)$, where $R$ is the radius of the high-density cloud. This implies a critical density for which Ly$\alpha$ photons do not escape ($f_\text{esc}^{2\gamma} \ll 1$) of
\begin{equation}
  n_\text{H,crit} \approx 10^6\,\text{cm}^{-3}\,T_4^{1/4}\left(\frac{x_e(1-x_e)}{10^{-4}}\frac{R}{0.3\,\text{pc}}\right)^{-1/2} \, .
\end{equation}

  \begin{figure}
    \centering
    \includegraphics[width=\columnwidth]{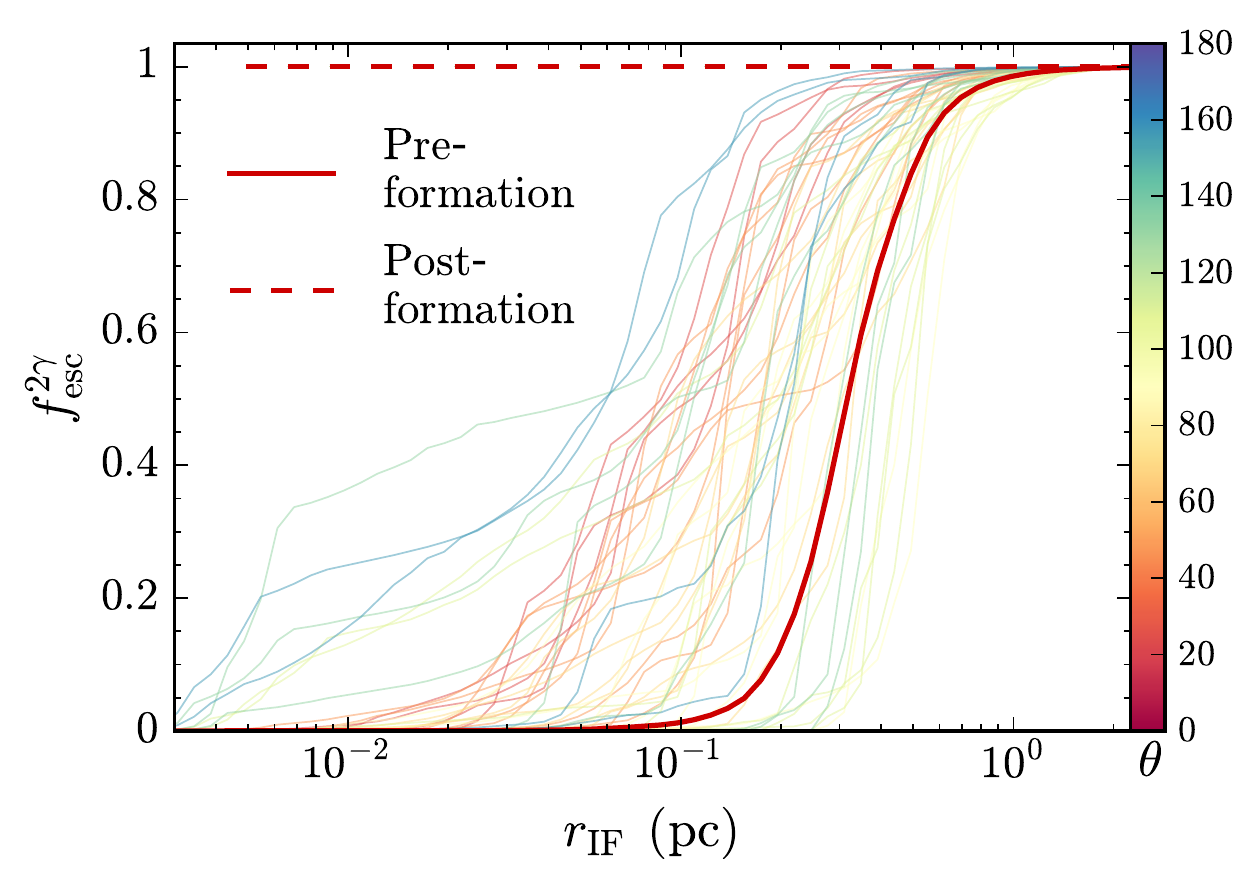}
    \caption{\protect\input{figures/f_esc_2p/caption}}
    \label{fig:f_esc_2p}
  \end{figure}

We also calculate $f_\text{esc}^{2\gamma}$ directly from the simulation data accounting for non-uniform conditions by taking the product from successive shells. The radial dependence is given by cumulative product within the region of interest
\begin{equation} \label{eq:f_esc_2p}
  f_\text{esc}^{2\gamma} = \prod \left[ 1 - p_\text{dest} \right]^{\Delta N_\text{scat}} \, ,
\end{equation}
where $\Delta N_\text{scat} \approx 0.6\,\Delta\tau_0$ denotes the contribution of each shell to the total number of scatterings. This shell-by-shell approximation is appropriate because additional scattering events take place under volume-weighted average conditions. The continuous version of equation~(\ref{eq:f_esc_2p}) is derived from the geometric product integral
\begin{align}
  f_\text{esc}^{2\gamma} &= \exp\left[\int \ln(1-p_\text{dest})\,\text{d}N_\text{scat} \right] \notag \\
    &\approx \exp\left[-0.6 \int p_\text{dest}\,n_{\HI}\,\sigma_0\,\text{d}r \right] \notag \\
    &\approx \exp\left[-\frac{3.1}{\text{pc}} \int \frac{x_e (1-x_e)}{10^{-4}} \left(\frac{n_\text{H}}{10^6\,\text{cm}^{-3}}\right)^2 T_4^{-1/2} \text{d}r \right] \, .
\end{align}
The exponential behaviour leads to a sharp cut-off corresponding to a characteristic length-scale for destruction of Ly$\alpha$ photons via two-photon decay. The degree of ionization also affects the likelihood of escape. Fig.~\ref{fig:f_esc_2p} shows the effect of a central ionizing source on the total $f_\text{esc}^{2\gamma}$, i.e. $r \gg r_\text{vir}$, as calculated from equation~(\ref{eq:f_esc_2p}). We assume a constant ionization fraction of $x_e \approx 10^{-4}$ within the ionization front. Specifically, for $r < r_\text{IF}$ the number densities from the hydrodynamical simulation are modified as follows: $n_p \approx n_\text{H}$ and $n_\text{\HI} \approx x_e n_\text{H}$ in the calculations for $p_\text{dest}$ and $\Delta N_\text{scat}$, respectively. We have checked that the result is not sensitive to the exact value of $x_e$. If the core remains neutral then essentially no Ly$\alpha$ photons are able to escape the core. However, as the DCBH forms and ionizes the gas, collisional de-excitation no longer regulates the Ly$\alpha$ luminosity, even for central emission. Finally, we note that Ly$\alpha$ radiation pressure may still be an important source of feedback in such compact environments despite the reduced force multiplication due to two-photon decay.

\subsection{Intrinsic luminosity}
During DCBH formation, excess thermal energy from the collapsing gas is efficiently radiated away via atomic line cooling with approximately $f_\alpha \sim 40\%$ of the release in gravitational binding energy being converted to Ly$\alpha$ emission \citep{Dijkstra_2014}. The gravitational potential energy is given by $\text{d}U = -G M_{<r} \text{d}m / r$, where the differential mass in spherical geometry is $\text{d}m = 4 \upi r^2 \rho\,\text{d}r$. The radial dependence of the intrinsic Ly$\alpha$ luminosity then is
\begin{equation}
  \frac{\text{d}L_\alpha^\text{grav}}{\text{d}r} = f_\alpha \frac{\text{d}\dot{U}}{\text{d}r} = 4 \upi f_\alpha G r \rho \dot{M} \, ,
\end{equation}
and the cumulative intrinsic Ly$\alpha$ luminosity is
\begin{equation}
  L_\alpha^\text{grav} = 16 \upi^2 f_\alpha G \int_0^R r^3 \rho^2 v_r\,\text{d}r \, .
\end{equation}
In practice, we calculate the luminosity as the cumulative sum from discrete cells or shells according to
\begin{equation} \label{eq:L_alpha_grav}
  L_\alpha^\text{grav} = f_\alpha \sum \frac{G\dot{M}}{r} \Delta m \, ,
\end{equation}
where gas must be infalling to contribute. The effects of dark matter are implicit in the gas density and velocity profiles. We note that Ly$\alpha$ cooling radiation arises from any physical mechanism that facilitates collisional excitation, collisional ionization, or recombination, e.g. shock heating, microturbulence, background radiation, etc. Such processes are indirectly included in our calculation of $L_\alpha^\text{grav}$ based on energy considerations and may be ignored once Ly$\alpha$ emission is dominated by black hole accretion and nebular sources.

  \begin{figure}
    \centering
    \includegraphics[width=\columnwidth]{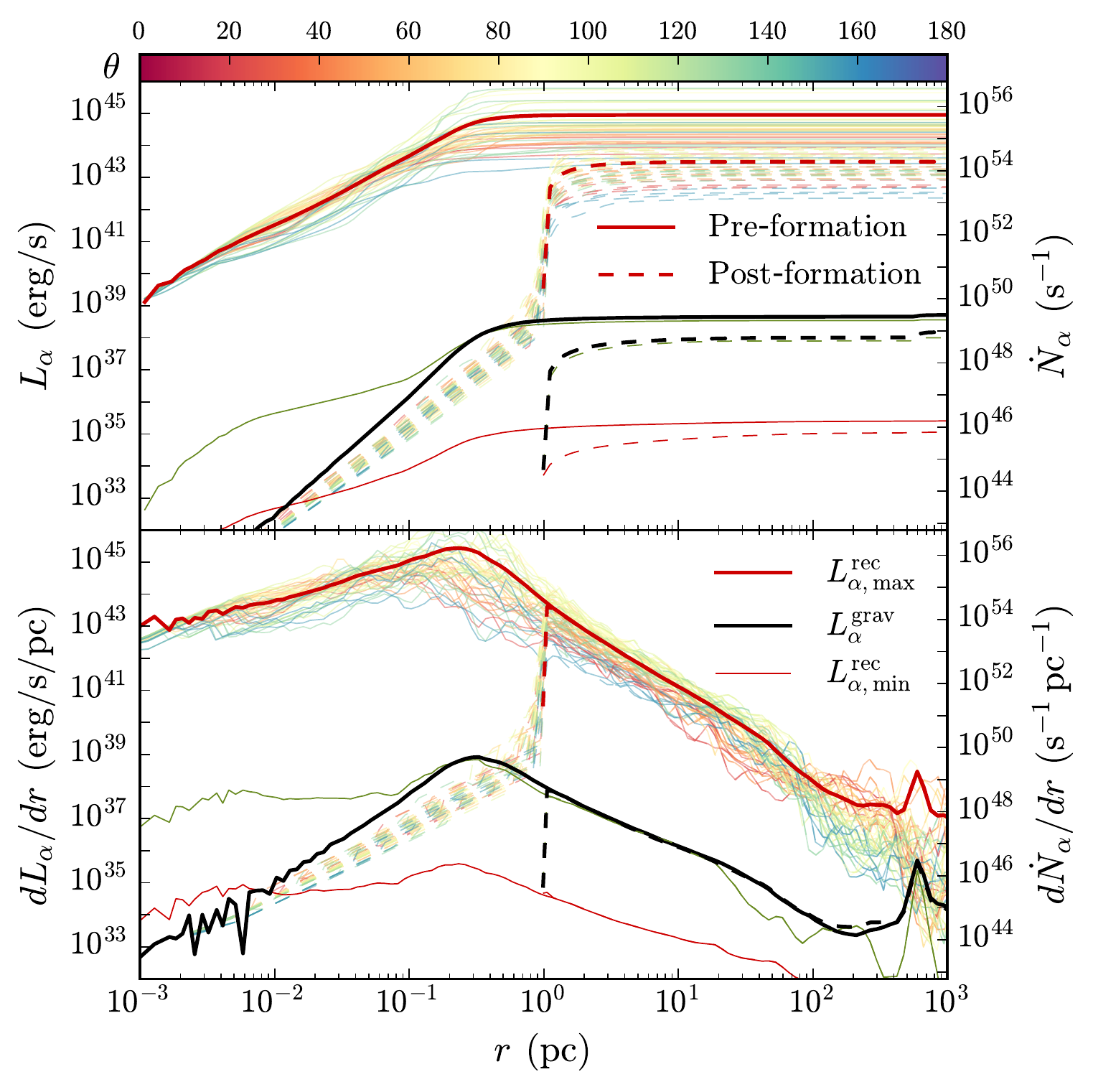}
    \caption{\protect\input{figures/L_r/caption}}
    \label{fig:L_r}
  \end{figure}

After the formation of a collapsed central object, ionizing radiation from the accretion disc leads to the efficient production of Ly$\alpha$ photons via recombinations. The cumulative intrinsic Ly$\alpha$ luminosity due to recombinations is
\begin{equation} \label{eq:L_alpha_rec}
  L_\alpha^\text{rec} = 0.68\,h\nu_\alpha \int_0^R 4 \upi r^2 \alpha_\text{B} n_e n_p \text{d}r \, ,
\end{equation}
where $h \nu_\alpha = 10.2$\,eV and the number densities $n_e$ and $n_p$ are for free electrons and protons \citep{Dijkstra_2014}. Fig.~\ref{fig:L_r} illustrates the radial dependence of Ly$\alpha$ emission from gravitational collapse and recombinations. The peak of $\text{d}L_\alpha/\text{d}r$ corresponds to the transition out of the core of the density profile, i.e. the break in the power law. In the pre-formation phase Ly$\alpha$ cooling dominates at $L_\alpha^\text{grav} \approx 4.4 \times 10^{38}\,\text{erg\,s}^{-1}$ while in post-formation recombinations from a central ionizing source can contribute as much as $L_\alpha^\text{rec} \sim 10^{45}\,\text{erg\,s}^{-1}$ (see Table~\ref{tab:L_scalings} for additional details).

\begin{table}
  \caption{Radial scaling relations for the Ly$\alpha$ luminosity calculated by equations~(\ref{eq:L_alpha_grav}) and (\ref{eq:L_alpha_rec}). The notation for the broken power law is the same as Table~\ref{tab:scalings}, i.e. $r_\textsc{i} \equiv r / (0.05~\text{pc})$ and $r_\textsc{ii} \equiv r / (10~\text{pc})$. Calculations in fully ionized and mostly neutral gas are denoted by the subscripts `max' and `min', respectively. As in Table~\ref{tab:scalings}, the scalings are given as pre-formation profiles in order to retain information on sub-parsec scales. However, the pre-formation case cannot be ionized so the cumulative value serves as an upper limit. A more conservative value for the luminosity after ionization breakout is $L_{\alpha,\text{max}}^\text{rec} \approx 5 \times 10^{43}\,\text{erg\,s}^{-1}$, corresponding to the ionized post-formation curve shown in Fig.~\ref{fig:L_r}.}
  \label{tab:L_scalings}
  \begin{tabular}{@{} lr ccc @{}}
    \hline
    Quantity & Units & Region \textsc{i} & Region \textsc{ii} & \hspace{-.525cm} $r_{\textsc{i}\text{-}\textsc{ii}}$\;[pc] \\
    \hline
    \vspace{.05cm}
    $L_{\alpha,\text{max}}^\text{rec}$ & \hspace{-.4cm} [erg/s] & \hspace{-.15cm} $1.16 \times 10^{43} \, r_\textsc{i}^{2.2}$ & \hspace{-.25cm} $\approx 9.0 \times 10^{44}$ & \hspace{-.15cm} $0.36$ \\
    $L_\alpha^\text{grav}$ & \hspace{-.4cm} [erg/s] & \hspace{-.15cm} $1.29 \times 10^{35} \, r_\textsc{i}^{3.9}$ & \hspace{-.25cm} $\approx 4.4 \times 10^{38}$ & \hspace{-.15cm} $0.40$ \\
    $L_{\alpha,\text{min}}^\text{rec}$ & \hspace{-.4cm} [erg/s] & \hspace{-.15cm} $3.18 \times 10^{33} \, r_\textsc{i}^{1.6}$ & \hspace{-.25cm} $\approx 3.0 \times 10^{35}$ & \hspace{-.15cm} $0.36$ \\
    $\frac{\text{d}}{\text{d}r}L_{\alpha,\text{max}}^\text{rec}$ & \hspace{-.4cm} [erg/s/pc] & $4.70 \times 10^{44} \, r_\textsc{i}^{1.2}$ & $1.33 \times 10^{41} \, r_\textsc{ii}^{-8/3}$ & \hspace{-.15cm} $0.23$ \\
    $\frac{\text{d}}{\text{d}r}L_\alpha^\text{grav}$ & \hspace{-.4cm} [erg/s/pc] & $7.91 \times 10^{36} \, r_\textsc{i}^{2.9}$ & $1.40 \times 10^{36} \, r_\textsc{ii}^{-2}$ & \hspace{-.15cm} $0.31$ \\
    $\frac{\text{d}}{\text{d}r}L_{\alpha,\text{min}}^\text{rec}$ & \hspace{-.4cm} [erg/s/pc] & $8.76 \times 10^{34} \, r_\textsc{i}^{0.6}$ & $1.85 \times 10^{33} \, r_\textsc{ii}^{-4/3}$ & \hspace{-.15cm} $0.26$ \\
    \hline
  \end{tabular}
\end{table}

However, the Ly$\alpha$ luminosity is also limited by the strength of the ionizing source. For concreteness, we relate the Ly$\alpha$ luminosity to the ionizing luminosity directly as
\begin{equation} \label{eq:L_ion}
  L_\alpha = \frac{3}{4} f_\text{coll} \left( 1 - f_\text{esc}^\text{ion} \right) L_\text{ion} \, ,
\end{equation}
which accounts for the fact that harder ionizing spectra can boost the overall Ly$\alpha$ production per ionizing photon \citep{Raiter_2010,Dijkstra_2014}. Here the ionizing luminosity is $L_\text{ion} \equiv \langle h \nu \rangle_\text{ion} \dot{N}_\text{ion} = \int_{\nu_\text{min}}^\infty L_\nu \text{d}\nu$ for a given source with $\nu_\text{min} = 13.6\,\text{eV}$. For reference, a blackbody source with an effective temperature of $T_\text{eff} = \{10^4, 10^5, 10^6\}\,\text{K}$ has a mean energy per ionizing photon of $\langle h \nu \rangle_\text{ion} = \{14.57, 29.61, 233.9\}\,\text{eV}$. The collisional parameter is defined as $f_\text{coll} \equiv (1 + 1.62\,n_\text{H,3}) / (1.56 + 1.78\,n_\text{H,3})$, where $n_\text{H,3} \equiv n_\text{H} / (10^3\,\text{cm}^{-3})$ and is bounded by $0.64 < f_\text{coll} < 0.91$ in the low- and high-density limits, respectively. The factor of $(1 - f_\text{esc}^\text{ion})$ accounts for the escape of ionizing photons either from 3D effects or because the galaxy is ionized out into the intergalactic medium (IGM). If we combine equations~(\ref{eq:Ndot_ion_BB}) and (\ref{eq:L_ion}) considering a blackbody with $T_\text{eff} = 10^5\,\text{K}$ emitting at the Eddington luminosity then the Ly$\alpha$ luminosity in the high-density, ionization-bound regime ($f_\text{coll} \approx 0.91$ and $f_\text{esc}^\text{ion} \ll 1$) is
\begin{align}
  L_\alpha &= f_\text{coll} \left( 1 - f_\text{esc}^\text{ion} \right) \frac{3\pi L_\text{Edd}}{4\sigma_\text{SB} T_\text{eff}^4} \int_{\nu_\text{min}}^\infty B_\nu\,\text{d}\nu \notag \\
                      &\approx 7.67 \times 10^{43}\,\text{erg\;s}^{-1}\,\left( \frac{M_\bullet}{10^6\,\Msun} \right) \, ,
\end{align}
or equivalently a rate of $\dot{N}_\alpha \approx 4.69 \times 10^{54}\,\text{s}^{-1}\,(M_\bullet / 10^6\,\Msun)$. In this case, the standard factor of $0.68$ Ly$\alpha$ photons per ionizing photon is increased to $f_\text{coll} \langle h \nu \rangle_\text{ion} / (13.6\,\text{eV}) \approx 2$. For comparison, higher effective temperatures or harder SEDs asymptotically approach the maximal Eddington-limited values of $L_{\alpha,\text{Edd}} \approx 8.58 \times 10^{43}\,\text{erg\;s}^{-1}\,(M_\bullet/10^6\,\Msun)$ and $\dot{N}_{\alpha, \text{Edd}} \approx 5.25 \times 10^{54}\,\text{s}^{-1}\,(M_\bullet / 10^6\,\Msun)$.

We note that such efficient conversion to Ly$\alpha$ emission assumes a low escape fraction, which may not be the case for long mean free path high energy photons. A more self-consistent calculation would perform multi-wavelength ionizing radiative transfer directly from the SED, which will be done in future 3D RHD simulations of DCBH growth. Still, X-ray feedback may have important consequences for Ly$\alpha$ transport, which is sensitive to the ionization state and dynamical evolution of the gas. For example, in neutral gas a harder spectrum leads to a higher residual ionized fraction due to X-ray heating \citep{Dijkstra_DCBH_2016}, while in ionized gas the longer mean free path implies a higher residual neutral fraction, which can increase the efficiency of Ly$\alpha$ driven winds \citep{Smith_CR7_2016}. In the following subsection we account for some of the uncertainty regarding the ionizing source by varying $x_\text{\HI} \equiv n_\text{\HI} / n_\text{H}$. Furthermore, the post-processing analysis allows for luminosity rescaling given a particular SED. Therefore, we may meaningfully refer to both direct emission properties, such as $\dot{N}_\alpha$, and model dependent quantities, such as $M_\bullet$.

\subsection{\texorpdfstring{Ly$\balpha$}{Lyα} radiative transfer simulations}
The MCRT post-processing methodology that enables accurate Ly$\alpha$ radiation pressure calculations with the Cosmic Ly$\alpha$ Transfer code (\colt) is described in detail by \citet{Smith_2015,Smith_RHD_2017}. In particular, we employ Monte Carlo estimators based on the traversed opacity for continuous momentum deposition to reduce noise in regions with fewer scatterings. In order to maintain the spatial resolution of the simulation, we have extended the capabilities of \colt\ to run natively on unstructured mesh data, i.e. the Voronoi tessellation of points. In fact, \colt\ is able to perform Ly$\alpha$ radiative transfer with any particle-based data under the interpretation of a Voronoi tessellation, which can be efficiently and robustly constructed with preexisting software such as the Computational Geometry Algorithms Library \citep{cgal_2017}. Implementing ray tracing and other algorithms in this geometry follows standard techniques \citep[e.g.][]{Springel_2010}, which have been carefully tested and optimized for Ly$\alpha$ radiative transfer.

  \begin{figure}
    \centering
    \includegraphics[width=\columnwidth]{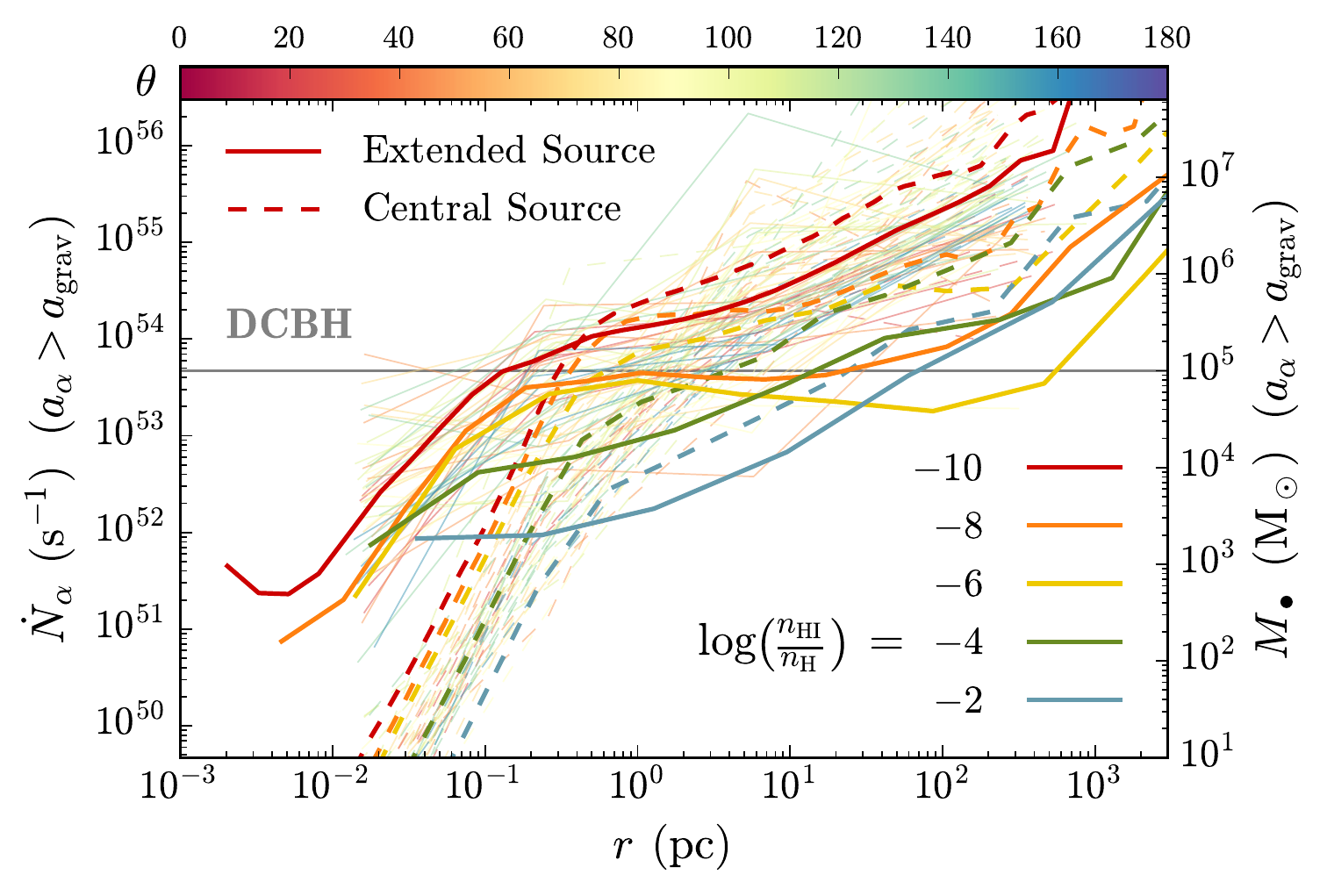}
    \caption{\protect\input{figures/a_ratio/caption}}
    \label{fig:a_ratio}
  \end{figure}

\subsubsection{Dynamical impact}
Ly$\alpha$ radiation pressure may contribute alongside other forms of feedback to impose restrictions on the maximum mass of the seed black hole. If Ly$\alpha$ feedback is able to overcome gravity and the momentum of the inflowing gas, then the resulting outflow could signal an end to the assembly process. In Fig.~\ref{fig:a_ratio}, we illustrate the relative dynamical importance of Ly$\alpha$ radiation pressure compared to gravity. In the limit of a central, compact source the Ly$\alpha$ radiation field can significantly overwhelm gravity within the core. Specifically, in the notation of Table~\ref{tab:scalings} we find the radial acceleration is $a_\alpha \sim \{10^{6\text{--}8}\,r_\textsc{i}^{-2.5}, 10^{1\text{--}3}\,r_\textsc{ii}^{-1.8}\}\,\text{km}\,\text{s}^{-1}\text{Myr}^{-1}\,(M_\bullet / 10^6\,\Msun)$ in regions~\textsc{i} and \textsc{ii} connecting at $r_{\textsc{i}\text{-}\textsc{ii}} \sim 0.5$\,pc, with the coefficient range arising from different ionization fractions. However, for an extended source distributed according to the recombination rate (see $L_\alpha^\text{rec}$ in Fig.~\ref{fig:L_r}), Ly$\alpha$ trapping is reduced in the core, while still having an impact. Interestingly, extended emission actually enhances the force outside the core because the Ly$\alpha$ photons have not diffused as far into the wings as in the central, bottled up scenario. For this same reason, we predict post-formation scenarios with reduced density in the core may allow Ly$\alpha$ radiation pressure to act on greater distances for similar luminosities, although we do not simulate this here. Overall, we expect Ly$\alpha$ feedback and gas pressure to contribute to the formation of galactic winds in response to the newly formed DCBH, given the typical mass range is $M_\bullet \sim 10^{4-6}\,\Msun$. In the case that the core remains ionization bounded or two-photon decay destroys Ly$\alpha$ photons, the extent of Ly$\alpha$ feedback will be limited to within the core. However, considering the low threshold to affect the gas in these regions Ly$\alpha$ trapping should eventually be explored in fully coupled 3D RHD simulations. Despite the considerable uncertainty in the interpretation of our results, the post-processing analysis is indicative of the complexity and computational requirements to fully determine the dynamical impact of Ly$\alpha$ radiation pressure.

  \begin{figure}
    \centering
    \includegraphics[width=\columnwidth]{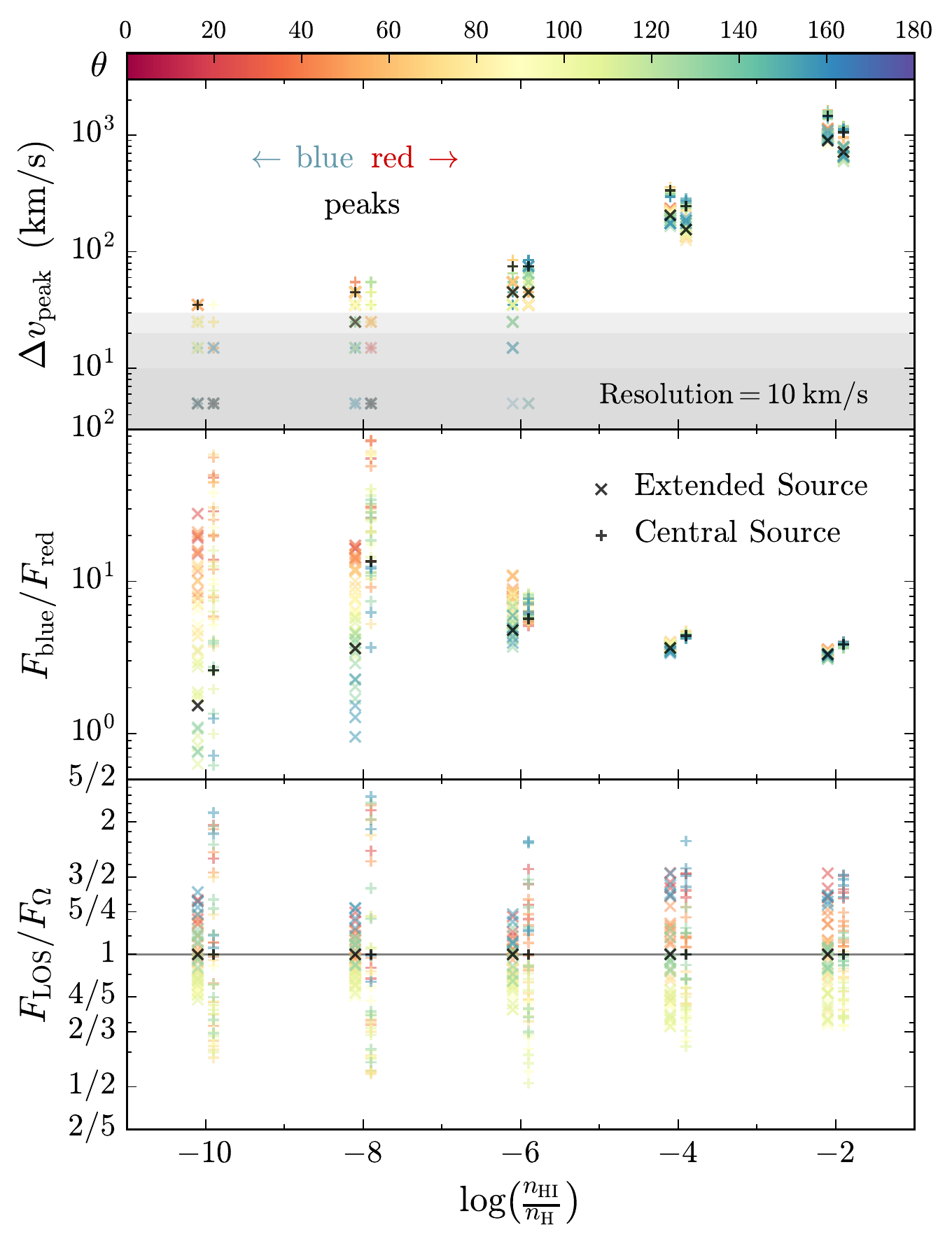}
    \caption{\protect\input{figures/Delta_v/caption}}
    \label{fig:Delta_v}
  \end{figure}

  \begin{figure}
    \centering
    \includegraphics[width=\columnwidth]{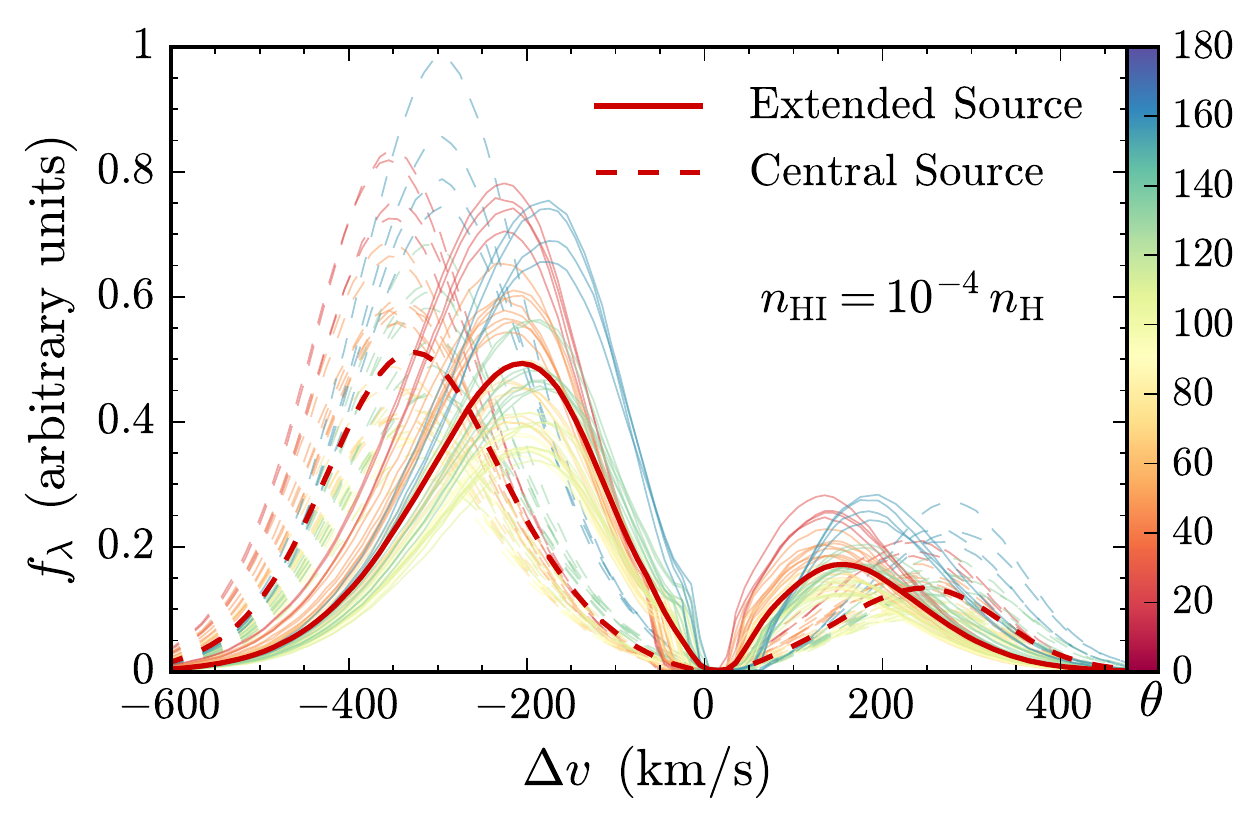}
    \caption{\protect\input{figures/Flux/caption}}
    \label{fig:Flux}
  \end{figure}

\subsubsection{Escape properties}
We now consider the characteristics of the intrinsic escape of Ly$\alpha$ photons during DCBH assembly. A single scattering with neutral hydrogen far from the source removes Ly$\alpha$ photons from the line of sight, so to first order all flux blueward of $\Delta v \lesssim 100\,\text{km\,s}^{-1}$ is eliminated. However, for simplicity we do not extrapolate to observed signatures based on IGM transmission. Instead we focus on the effects of the 3D geometry. Fig.~\ref{fig:Delta_v} shows three basic parameters of the emergent line profiles for different values of $x_\text{\HI} \equiv n_\text{\HI} / n_\text{H}$ with either central or extended emission.

(i) The top panel demonstrates the strong effect of residual opacity on the velocity offset of the blue and red peaks, $\Delta v_\text{peak}$. Depending on the ionization state, the peaks are likely to emerge with a velocity offset and FWHM of $\sim 10^{2-3}\,\text{km\,s}^{-1}$. Another interesting feature is the angular dependence of the peak location, dominated by a dipole-like Doppler shift of $\Delta(\Delta v_\text{peak}) \approx \pm v_\text{sys} \approx \pm 24\,\text{km\,s}^{-1}$ due to the systemic velocity of the source with respect to the centre of mass, in addition to random velocity dispersions. Overall, the blue peak has a larger absolute velocity offset than the red peak, i.e. $|\Delta v_\text{blue}| > |\Delta v_\text{red}|$, and the peak separation has little variation across sightlines, i.e. $|\Delta v_\text{blue}| + |\Delta v_\text{red}| \approx \text{constant}$.

(ii) The middle panel shows the directional dependence of the blue to red flux ratio, $F_\text{blue} / F_\text{red}$. In the collapsing scenario, the blue peak dominates by a factor of a few with significant angular dependence in the optically thin limit, while converging to a factor of a few in the high-opacity limit. We also note that an extended source results in a more balanced profile (lower ratio) than a central source.

(iii) The lower panel illustrates the line-of-sight bolometric Ly$\alpha$ flux, $F_\text{LOS}$, compared to the isotropic case of $F_\Omega \equiv L_\alpha / (4 \upi d_\text{L}^2)$, where $d_\text{L}$ denotes the luminosity distance. Due to the ellipsoidal or thick-disc geometry \citep[e.g.][]{Lodato_2006}, we find beaming along both directions of the angular momentum axis by a factor of a few compared to edge-on sightlines. It is important to note that the excess is less pronounced for extended emission than for a central source, although the difference may not matter in the optically thick limit.

Additional insights regarding the escape of Ly$\alpha$ photons in 3D DCBH environments are available upon close examination of the line profile of a single simulation. We choose the case of $x_\text{\HI} = 10^{-4}$ and show the angular-dependence of the emergent flux density and bolometric flux in Fig.~\ref{fig:Flux}. All of the profiles show an enhancement of the blue peak due to gas infall, such that $F_\text{blue} / F_\text{red} \approx 4$. The line-of-sight flux is also noticeably enhanced by at least a factor of two along either direction of the rotation axis. Also, for a given simulation the locations of the blue and red peaks varies by approximately $50\,\text{km\,s}^{-1}$ across different sightlines due to a systemic dipole velocity shift. The absolute peak separation is about $1.5$ times smaller for the extended emission compared to the central source due to the lower effective optical depth. The shape of the profile is also noticeably different because each photon is emitted in distinct regions with unique pathways to escape. Interestingly, the more realistic recombination emission setup produces a more skewed or less symmetric line, similar to the observed profiles of many Ly$\alpha$ emitting galaxies \citep[e.g.][]{Rivera-Thorsen_2015}.

\section{Compton Scattering}
\label{sec:compton}

\subsection{Initial stage}
Before the \HII\ region develops, the number density of free electrons is too low for significant Compton scattering. We calculate a lower limit for the Thomson optical depth of $\tau_\text{T} = N_e \sigma_\text{T} \gtrsim 10^{-5}$. Therefore, if the DCBH environment becomes optically thick to Compton scattering it is in response to ionizing radiation emitted from the newly formed black hole. However, as discussed previously, we do expect the gas to become ionized so the opacity calculated from the pre- and post-formation electron column densities will rise and has the potential to eventually rise to $\tau_\text{T} \gtrsim 1$ with the growth of the DCBH, rendering the emitting region Compton-thick.

\subsection{Upper limits}
\begin{figure}
    \centering
    \includegraphics[width=\columnwidth]{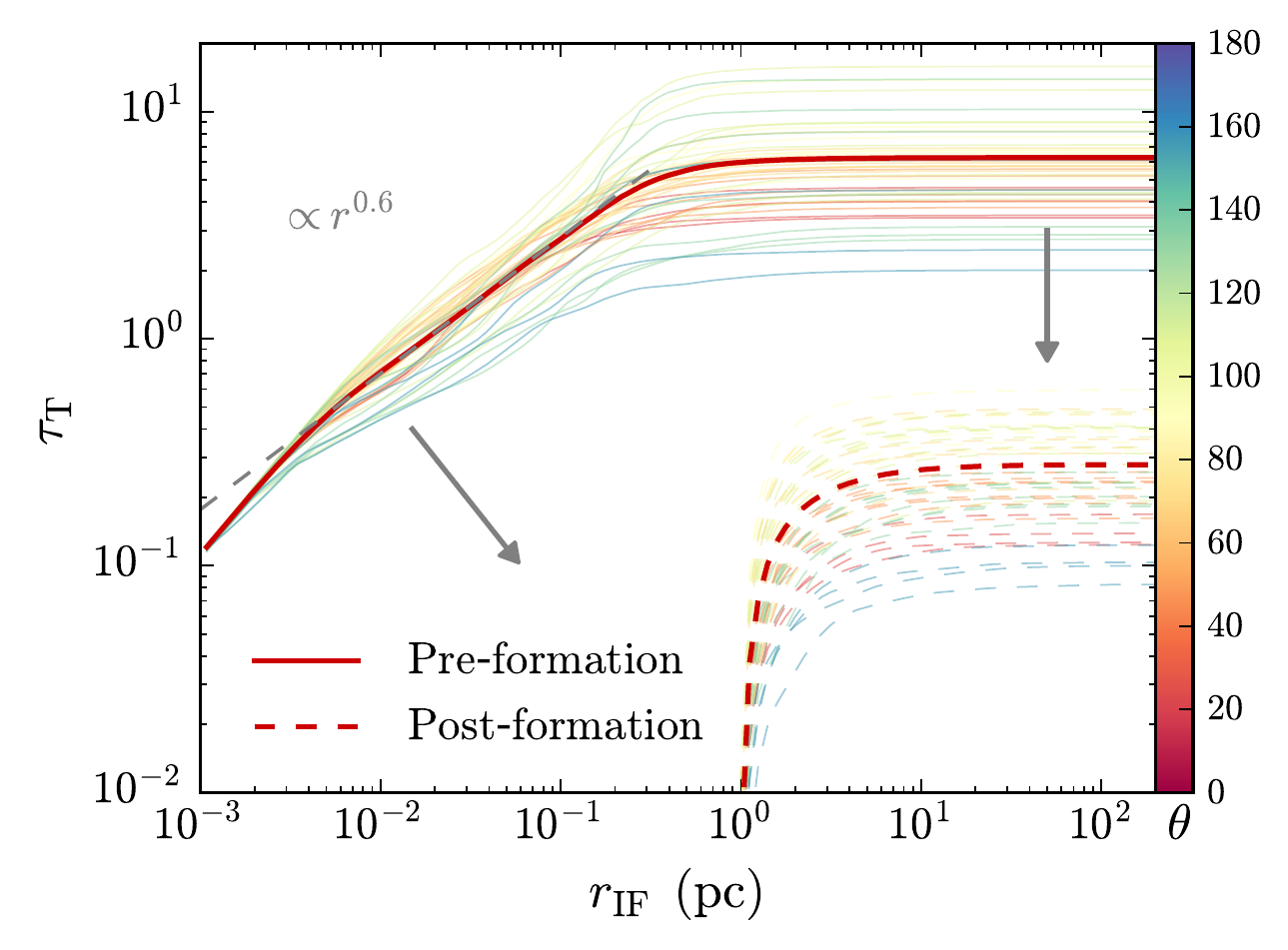}
    \caption{\protect\input{figures/tau_T/caption}}
    \label{fig:tau_T}
  \end{figure}

Fig.~\ref{fig:tau_T} illustrates the effect of the ionization front radius $r_\text{IF}$ on the cumulative optical depth to Compton scattering. Although we do not follow the hydrodynamical evolution beyond the initial formation, our extrapolation indicates that DCBH environments become Compton-thick for a significant fraction of sightlines. 1D RHD simulations combined with spectral synthesis modelling have already been carried out by \citet{Pacucci_MBH_Spectra_2015} \citep[see also][]{Yue_2013}. They found that Compton-thickness does indeed occur, persisting for about $120$\,Myr under their standard accretion scenario. Furthermore, the emerging spectrum is strongly affected by the hydrogen column density as photons shortward of the Ly$\alpha$ line are reprocessed to lower energies. Thus, the predicted spectral signature of DCBHs exhibits characteristic non-thermal emission in the observed infrared and X-ray bands. The conditions for Compton-thickness still need to be explored with 3D ab initio simulations, including self-consistent buildup of the \HII\ region and all relevant forms of radiative feedback.

An additional consideration is whether ongoing accretion is able to replenish the gas supply after the rapid formation of a central massive object. If runaway collapse efficiently evacuates the central parsec region, the replacement time is approximately $t_\text{acc} \approx M_{<r} / \dot{M} \gtrsim 72.3\,\text{kyr}\,(r/\text{pc})^{4/3}$, which is longer than the dynamical time by a factor of $t_\text{acc} / t_\text{ff} \gtrsim 1.45\,(r/\text{pc})^{1/6}$. However, an extrapolation in time suggests that gas is not flowing into the core fast enough to maintain Compton-thickness indefinitely. At some point, the combination of decreasing column density and increasing radiation pressure dramatically alters the environment, leading to sub-Eddington black hole growth.

We emphasize that rapid evacuation of the central gas during the assembly of the initial seed black hole may reduce the post-formation accretion rates. For example, \citet{Pacucci_2015} suggest that super-Eddington accretion can occur for extended periods ($\gtrsim 100$\,Myr), however, a lower ambient density could imply an earlier transition to feedback-limited growth rates. On the other hand, the authors also point out that a radiatively inefficient slim disc mode of accretion could allow for sub-Eddington luminosities and super-Eddington rates. Either way, 3D radiation hydrodynamics with ionizing and Ly$\alpha$ feedback in (near) Compton-thick settings may moderate extreme hyper-Eddington accretion in the aftermath of direct collapse.

\section{Summary and Conclusions}
\label{sec:conc}
We have explored the impact of radiative feedback on the assembly environments of direct collapse black holes. The main focus has been to apply a Monte Carlo Ly$\alpha$ radiative transfer code to calculate the pressure exerted by Ly$\alpha$ photons on the surrounding gas. Previous applications of this method have been limited to 1D geometries, either in the context of the expanding shell model \citep{Dijkstra_Loeb_2008}, or coupled to a spherically symmetric hydrodynamics code \citep{Smith_RHD_2017}. In this work, we have performed an exploratory post-processing analysis of the high-resolution ab initio cosmological simulation of \citet{Becerra_2017}. We found that Ly$\alpha$ radiation pressure \textit{is} likely important in shaping the environment surrounding newly formed DCBHs. Therefore, 3D Ly$\alpha$ radiation hydrodynamics will be crucial to incorporate in future DCBH simulations, alongside accurate ionizing radiative transfer and Compton scattering as discussed above.

Our results confirm previous predictions that Ly$\alpha$ radiation can have a dynamical impact in realistic astrophysical settings, beyond its role as a probe of (inter)stellar and (inter)galactic properties. Still, determining the exact role of Ly$\alpha$ photon trapping is challenging both theoretically and observationally. For example, the Monte Carlo method is computationally demanding in extremely optically thick environments and great care must be taken to ensure convergence at every stage in fully coupled RHD simulations. However, more efficient algorithms and hardware will eventually make such calculations feasible. Additionally, upcoming observations with the \textit{JWST} and other facilities extending our view of the high-redshift frontier will help guide and motivate simulations to better understand newly discovered phenomena. Still, our models robustly predict intense Ly$\alpha$ radiation pressure within the sub-parsec region. Certain configurations may extend the influence of Ly$\alpha$ feedback out to $\sim 100$\,pc, although the quantitative details are likely sensitive to a number of environmental factors including photoionization, two-photon decay, velocity gradients, dust absorption and geometrical effects.

Details regarding the impact of Ly$\alpha$ feedback in the direct collapse scenario will remain uncertain until fully cosmological, on-the-fly simulations are available. Still, Ly$\alpha$ trapping likely induces thermal effects via less efficient cooling, chemical effects via photodetachment or photodissociation, and kinematic effects via direct momentum transfer. The thermal impact has recently been revisited by \citet{Ge_Wise_2017}, who perform time-dependent Ly$\alpha$ radiative transfer on radially averaged profiles during the initial collapse of massive black hole seeds. Similar to \citet{Spaans_Silk_2006}, the equation of state for gas at moderate densities ($n \sim 10^4$ -- $10^5$) could potentially have $\gamma$ as high as 4/3, which would briefly affect the Jeans mass and reduce fragmentation. \citet{Ge_Wise_2017} claim a 50\,000\;K envelope forms at a radius of $\sim 1$\;pc. However, the core still maintains near-isothermal collapse as cooling proceeds via other atomic hydrogen transitions and two-photon emission \citep[e.g.][]{Omukai_2001,Schleicher_2010}. Our work based on the simulation of \citet{Becerra_2017} already includes an approximate prescription for Ly$\alpha$ trapping in this phase. Thus, we focus our investigation on the longer-term dynamical impact of Ly$\alpha$ radiation pressure once the massive object has formed. The importance of Ly$\alpha$ feedback during DCBH assembly seems fairly robust, however, fully coupled 3D calculations may be necessary for a self-consistent, higher order description. Of course, fundamental insights about Ly$\alpha$ radiative transfer can guide our understanding. For example, a static cloud of constant density has an approximate temperature dependence for the trapping time of $t_\text{trap} /t_\text{light} \sim (a \tau_0)^{1/3} \propto T^{-1/3}$ \citep{Adams_1975}, and the characteristic escape frequency changes as $\Delta v_\text{peak} = x_\text{peak} v_\text{th} \propto T^{1/6}$, where $x_\text{peak} \sim (a \tau_0)^{1/3}$ and $v_\text{th} \propto T^{1/2}$ \citep[e.g.][]{Smith_2015}. Therefore, in the context of fully coupled simulations, we expect our results to be modified by other environmental factors, such as the rapid development of low-column pockets to facilitate the escape of Ly$\alpha$ photons.

Ly$\alpha$ trapping may also have an effect on other observational signatures. For example, additional radiation pressure provides negative feedback, making it more difficult for gas to reach the source, thereby regulating the black hole mass and luminosity $\propto \dot{M}_\bullet$. The emerging SED is also indirectly related to Ly$\alpha$ processes as additional feedback may destroy the Compton-thickness property and excess two-photon emission earlier than expected. Ly$\alpha$ photons may also affect chemical processes via the photodetachment of ions and molecules. Finally, Ly$\alpha$ spectroscopy with next-generation observatories may provide further insights into high-$z$ galaxies and radiative transfer effects, such as the presence (absence) of outflowing gas. In future studies it may be worth revisiting Ly$\alpha$ radiation pressure in the context of other environments as well. Ly$\alpha$ coupling is likely sub-dominant in simulations of massive star formation, including Population~III stars \citep{McKee_Tan_2008,Stacy_2012}, and galactic winds caused by AGN, starbursts and supernovae \citep{Haehnelt_1995}. However, there may be aspects in which these systems can still be dynamically influenced by Ly$\alpha$ trapping due to a radial scaling steeper than free-streaming, $\text{d}\log a_\alpha / \text{d}\log r < -2$, from force multiplication in the diffusion limit \citep{Dijkstra_Loeb_2008,Smith_RHD_2017}. In any case, Ly$\alpha$ photons are crucial in shaping conditions in the early Universe, and in allowing us to probe them.

\section*{Acknowledgements}
This material is based upon work supported by a National Science Foundation Graduate Research Fellowship for AS. VB acknowledges support from NSF grant AST-1413501. We thank Benny Tsang for insightful discussions. We thank the anonymous referee for the prompt and constructive comments that improved the content of this paper. The authors acknowledge the Texas Advanced Computing Center~(TACC) at the University of Texas at Austin for providing HPC resources.


\bibliographystyle{mnras}
\bibliography{biblio}

\bsp 
\label{lastpage}
\end{document}